\documentclass{emulateapj}

\usepackage{xcolor}

\shorttitle{Spatially Resolved Gas Kinematics within a \lya\ Nebula}
\shortauthors{Prescott et al.}

\newcommand{\apsizeMID}{7.2} 
\newcommand{\apsizeA}{5.8} 
\newcommand{\apsizeMIDkpc}{61} 
\newcommand{\apsizeAkpc}{49} 

\newcommand{\ergss}{erg~s$^{-1}$}
\newcommand{\degree}{\ensuremath{^\circ}}
\newcommand{\bw}{$B_{W}$}

\newcommand{\cii}{\hbox{{\rm C}\kern 0.1em{\sc ii}}}
\newcommand{\civ}{\hbox{{\rm C}\kern 0.1em{\sc iv}}}
\newcommand{\ciii}{\hbox{{\rm C}\kern 0.1em{\sc iii}}]}
\newcommand{\heii}{\hbox{{\rm He}\kern 0.1em{\sc ii}}}
\newcommand{\halpha}{\hbox{{\rm H}\kern 0.1em$\alpha$}}
\newcommand{\hbeta}{\hbox{{\rm H}\kern 0.1em$\beta$}}
\newcommand{\neiv}{\hbox{{\rm Ne}\kern 0.1em{\sc iv}}]}
\newcommand{\nv}{\hbox{{\rm N}\kern 0.1em{\sc v}}}
\newcommand{\oi}{[\hbox{{\rm O}\kern 0.1em{\sc i}}]}
\newcommand{\oii}{[\hbox{{\rm O}\kern 0.1em{\sc ii}}]}
\newcommand{\oiii}{[\hbox{{\rm O}\kern 0.1em{\sc iii}}]}

\newcommand{\lya}{Ly$\alpha$}

\begin{document}

\title{Spatially Resolved Gas Kinematics within a \lya\ Nebula: Evidence for Large-scale Rotation} 

\author{Moire K. M. Prescott\altaffilmark{1,2}, Crystal L. Martin\altaffilmark{2}, \& Arjun Dey\altaffilmark{3,4}}

\altaffiltext{1}{Dark Cosmology Centre, Niels Bohr Institute, University of Copenhagen, Juliane Maries Vej 30, 2100 Copenhagen Ø, Denmark; mkmprescott@dark-cosmology.dk}
\altaffiltext{2}{Department of Physics, Broida Hall, Mail Code 9530, University of California, Santa Barbara, CA 93106, USA} 
\altaffiltext{3}{National Optical Astronomy Observatory, 950 North Cherry Avenue, Tucson, AZ 85719, USA} 
\altaffiltext{4}{Radcliffe Institute for Advanced Study, Byerly Hall, Harvard University, 10 Garden Street, Cambridge, MA 02138, USA} 

\begin{abstract}

We use spatially extended measurements of \lya\ as well as less optically thick
emission lines from an $\approx80$~kpc \lya\ nebula at $z\approx1.67$ to assess the role of resonant scattering 
and to disentangle kinematic signatures from \lya\ radiative transfer effects.  We find 
that the \lya, \civ, \heii, and \ciii\ emission lines all tell a similar story in this system,
and that the kinematics are broadly consistent with large-scale rotation.
First, the observed surface brightness profiles are similar in 
extent in all four 
lines, strongly favoring a picture in which the 
\lya\ photons are produced in situ instead of being resonantly scattered from a central source.
Second, we see low kinematic offsets between \lya\ and
the less optically thick \heii\ line ($\sim100-200$ km s$^{-1}$), 
providing further support for the argument that the \lya\ and other 
emission lines are all being produced within the spatially extended gas.
Finally, the full velocity field of the system shows 
coherent velocity shear in all emission lines:
$\approx$500 km~s$^{-1}$ over the central $\approx$50~kpc of the nebula.
The kinematic profiles are broadly consistent with
large-scale rotation in a gas disk that is at least partially stable against collapse.
These observations suggest that the \lya\ nebula represents accreting material that is illuminated by an offset, 
hidden AGN or distributed star formation, and that is undergoing rotation in a clumpy and turbulent gas disk.  
With an implied mass of M($<R=20$~kpc)$\sim3\times10^{11}$ M$_{\odot}$, this system may represent 
the early formation of a large Milky Way mass galaxy or galaxy group.

\end{abstract}


\section{Introduction}
\label{sec:intro}

Giant Ly$\alpha$ nebulae (or ``\lya\ blobs'') 
are signposts of active galaxy formation.  
The most luminous examples, which exceed $\sim100$~kpc in size and 
$\sim10^{44}$ \ergss\ in \lya\ luminosity, are rare and found primarily 
in large-scale overdensities 
\citep[e.g.,][]{stei00,mat04,mat05,sai06,pres08,yang09,yang10,mat09,mat11,erb11,pres12a,pres13}.  
These spatially-extended gaseous nebulae often coexist with star-forming 
galaxies (i.e., \lya-emitting galaxies, Lyman break galaxies, 
submillimeter galaxies) and obscured AGN \citep[e.g.,][]{chap04,mat04,dey05}, suggesting these regions 
are galaxy groups or clusters in formation 
\citep[e.g.,][]{presphd,yang09,pres12b}.
Thus, giant \lya\ nebulae contain important clues to the dominant physical 
mechanisms at work during episodes of massive galaxy formation and offer an 
observational window into the flow and enrichment of gas within the cosmic web.

Many previous observational studies have focused on the potential power sources within 
\lya\ nebulae \citep[e.g.,][]{chap04,basu04,dey05,gea09,pres09,webb09,colbert11,stei11,yang11a,pres12b}, 
on their polarization properties \citep{pres11,hayes11}, on the properties of their dust and molecular gas 
\citep{chap01,yang12,yang14a}, or on their detailed morphology \citep{pres12b}.  
Theoretical studies have investigated a variety of powering scenarios: shock-heating in galactic superwinds 
\citep[e.g.,][]{tani00,tani01,mori04}, gravitational cooling in infalling cold streams 
\citep[e.g.,][]{hai00,far01,yang06,dijkloeb09,goerdt10,faucher10,rosdahl12}, and resonant scattering or photoionization due to emission from AGN or star formation \citep[e.g.,][]{cantalupo05,koll10,zheng11,cen13}. 
Two key issues have emerged from these studies of large \lya\ nebulae: 
(a) how much of the \lya\ emission in these systems is scattered over large spatial 
scales from a central source versus produced in situ within the nebula and 
(b) what are the underlying kinematics of the gas?

Theoretically, we know that \lya\ photons should be subject to substantial resonant scattering 
under typical astrophysical conditions.  
For the neutral hydrogen column densities typical of Lyman limit systems ($\sim10^{20}$ cm$^{-2}$), 
the optical depth at line center is of order $10^{7}$ \citep[e.g.][]{ver06}.
Resonant scattering leads to double-peaked emission line profiles as \lya\ photons must 
diffuse into the wings of the line before they are able to escape the system 
\citep[e.g.,][]{neu90,dijk06a,ver06}.  
The large number of scatterings also means that the dust content and its distribution within 
the system can have a profound effect on the emergent line, preferentially suppressing the \lya\ equivalent width in the 
case of a diffuse distribution or boosting it in the case of clumpy dust (\citealt{neu90,han06}, but also 
see \citealt{duval14,laursen2013}).  
At the same time, gas kinematics can profoundly alter the emission line profile of \lya\ as 
it emerges from the system, with outflows and infall leading to preferential absorption of the blue or 
red portion of the line, respectively \citep[e.g.,][]{dijk06a,ver06,laursen09a}.  
For sources at high redshift, absorption of the blue side of the \lya\ profile by the 
intergalactic medium also becomes important \citep{mad95}.  
All these effects are encoded in the shape of the \lya\ line profile, so 
deciphering them requires a side-by-side comparison to 
a non-resonant line, i.e., a tracer that is 
not susceptible to resonant scattering, but that is detected from the same region.

As the second brightest emission line of Hydrogen after \lya, the \halpha\ emission line is 
perhaps the best non-resonant comparison line for this type of study.  Unfortunately, 
most known \lya\ nebulae have been found at redshifts where \halpha\ is difficult to observe 
from the ground.  An important exception is a sample of 
\lya\ nebulae selected by \citet{yang10} at $z\approx2.3$.  
Using follow-up optical/NIR spectroscopy of \lya\ as well as non-resonant emission 
lines (e.g., \oiii, \halpha), they investigated the kinematics of eight 
of the larger \lya\ nebula systems from their sample \citep{yang11a,yang14b}.  
However, the ground-based NIR spectroscopic observations in these studies yielded high 
signal-to-noise ratio information only at the position of galaxies embedded within each \lya\ 
nebula system but were not used to investigate non-resonant line emission (if present) from 
the extended gas within the nebula itself.  
A similar analysis was carried out using \oiii\ detections from two embedded galaxies within 
a \lya\ nebula at $z\approx3.1$ \citep{mcl13}, while a recent IFU study detected \oii, \oiii, 
and \halpha\ within the central $\approx30$~kpc of a \lya\ nebula at $z=2.38$ \citep{overzier13}. 

Probing the kinematics and ionization structure of diffuse gas over larger spatial scales and 
in regions far from any associated bright galaxies requires mapping out emission-line nebulae in 
both \lya\ as well as less optically thick emission lines.
In this context, our discovery of a giant \lya\ nebula (henceforth referred to as PRG1) with bright, 
spatially-extended \heii$\lambda$1640 emission and somewhat weaker metal lines \citep{pres09,pres12a,pres13} 
provides a rare opportunity.  
\heii$\lambda1640$ (``\halpha'' for singly ionized Helium) is a non-resonant line but, unlike \halpha, it is 
located in the observed optical at these redshifts where ground-based observations are substantially more 
sensitive than in the NIR.  
This source therefore allows us to empirically investigate how much the observed \lya\ emission is affected 
by radiative transfer effects and study the kinematics and ionization of {\it the 
spatially extended gas} within a giant \lya\ nebula.

We report the observations and reductions in Section~\ref{sec:obsredux} and present the reduced spectra, 
as well as the surface brightness, emission line, and kinematic profiles in Section~\ref{sec:results}.  
In Section~\ref{sec:discussion} we discuss the implications of these results, and we conclude in 
Section~\ref{sec:conclusions}.  
A companion paper will use these same data to explore the energetics of the system as well as 
any spatial variations in the physical conditions (e.g., metallicity, density, ionization parameter) 
within the extended \lya\ nebula.  

In this paper, we assume the standard $\Lambda$CDM cosmology ($\Omega_{M}$=0.3, $\Omega_{\Lambda}$=0.7, $h$=0.7);
the angular scale at $z=1.67$ is 8.47 kpc/\arcsec.  All magnitudes are in the AB system \citep{oke74}.


\section{Observations \& Reductions}
\label{sec:obsredux}

We obtained deep spectroscopic observations of PRG1 using the Low Resolution Imaging Spectrometer 
\citep[LRIS;][]{oke95} on the Keck I Telescope during two separate nights on UT 2009 April 25 and UT 2010 May 10.  
A summary of the observations is given in Table~\ref{tab:specobs}.  
All observations used the 400/3400 grism on LRIS-Blue and the 400/8500 grating on LRIS-Red.  
For each run the data were taken in multislit mode to ensure accurate and repeatable pointing.  
Masks were designed with a long slitlet centered on PRG1 
and the remaining slitlets centered on filler targets in the vicinity.  
The position of the target on the slitmasks was adjusted to ensure LRIS-Blue wavelength coverage blueward 
of $\sim$3250\AA, the wavelength of \lya\ at $z\approx1.67$.  

For the 2009 run, 
the LRIS-Red grating angle was 
set to achieve full coverage from the D560 dichroic edge to $\sim$10,000\AA.  
The slitlets centered on PRG1 were $\approx36-44$\arcsec\ in length, 
the slitlet widths were set to 1\farcs2, 
and the data were unbinned.  
The resulting spectral resolution at the wavelengths of \lya\ and \heii\ was 
$\sigma\approx315$ and $215$ km~s$^{-1}$, respectively. 
For the 2010 run, 
we used the D680 dichroic and set the LRIS-Red grating angle to achieve full coverage from 
the dichroic edge out to $\sim$10,870\AA, in order to cover the \oii\ emission line at $z\approx1.67$.  
The slitlets centered on PRG1 were $\approx48-50$\arcsec\ in length, 
the slitlet widths were 1\farcs5, 
and the LRIS-Blue data were binned by 2 in the spectral direction in order to reduce the effect of readnoise 
and increase the signal-to-noise ratio.  
The resulting spectral resolution at the wavelengths of \lya\ and \heii\ was 
$\sigma\approx395$ and $317$ km~s$^{-1}$, respectively. 

\begin{figure}[h]
\includegraphics[angle=0,width=6in]{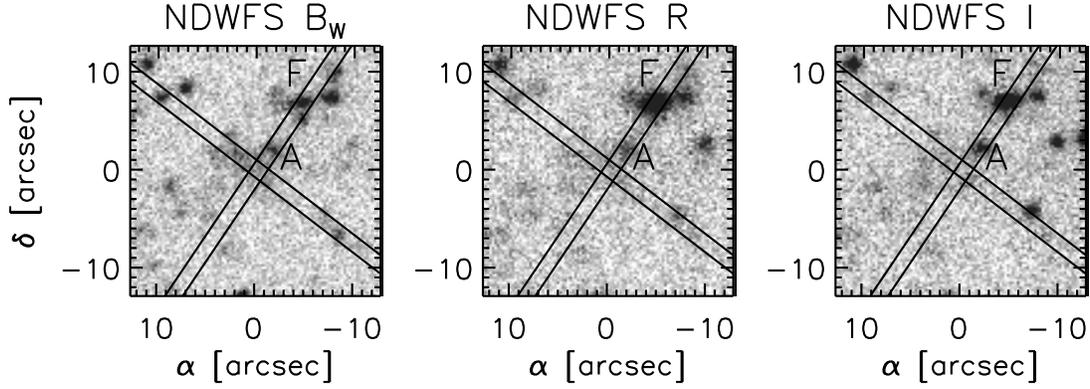} 
\caption[]{
Broad-band \bw, $R$, and $I$-band imaging of PRG1, oriented with N up and E to the left, overlaid with 
the two slit positions used for spectroscopic observations (PA=52\fdg44 and PA=146\fdg0).  
Diffuse continuum emission from the nebula is visible, particularly in the \bw\ image \citep{pres13}.  
In each panel, the origin is located at 14:35:12.385 +35:11:06.62, the 
location where the two slits cross, and Source `A' - the red, compact 
source at the northwest edge of the nebula - is labeled.  
Note that the source labeled `F,' the bright galaxy intersecting the PA=146\fdg0 
slit at [-5\arcsec,7\arcsec], is a $z\approx0.479$ object, with unambiguous 
\oii, \hbeta, and \oiii\ emission visible in the LRIS spectroscopy.
}
\label{fig:obssetup}
\end{figure}

During each run, the target was observed using two slit position angles: PA=52\fdg44, 
chosen to traverse the longest dimension of the diffuse emission seen in the ground-based \bw\ image, and PA=146\fdg0, 
chosen to cover both diffuse emission and the compact red source 
located at the northwest edge of the nebula \citep[Source A;][]{pres09}.  
The two slit positions are shown overlaid on the broad-band \bw, $R$, and $I$-band images of PRG1 in Figure~\ref{fig:obssetup}.  
While there was cirrus at sunset during the 2009 run, it cleared quickly 
and the observations were taken under clear conditions and 
0\farcs8-1\farcs1 seeing; 
during the 2010 run, the conditions were clear with 0\farcs7-0\farcs9 seeing.  
Individual exposures were dithered by $\approx4-7$\arcsec\ in 2009 and $\approx7$\arcsec\ in 2010.  
The total exposure times on LRIS-Blue/LRIS-Red were 5.2/4.8~hrs 
for PA=52\fdg44 
and 4.9/4.8~hrs 
for PA=146\fdg0  
during the 2009/2010 observing runs.

The data were reduced using IRAF and a customized version of the {\sc bogus} reduction package.\footnote{{\sc bogus} was 
written by Andrew J. Bunker, S. Adam Stanford, \& Daniel Stern: https://zwolfkinder.jpl.nasa.gov/~stern/homepage/bogus.html.}  
The first step was to apply the overscan correction and multiply the individual frames by the appropriate gain.  
Flatfield corrections were applied using domeflat and twilight exposures for the LRIS-Blue data and domeflat frames for the LRIS-Red data.  
Custom bad pixel masks were used to interpolate over bad columns and the script {\sc l.a.cosmic}\footnote{{\sc l.a.cosmic} 
was written by Pieter G. van Dokkum: http://www.astro.yale.edu/dokkum/lacosmic/.} was used to identify cosmic rays, both 
of which were particularly numerous in the 2010 LRIS-Red data.  The individual exposures were shifted and stacked to generate 
the final 2D spectra.  
The spectral trace was determined using brighter reference objects along the slit, and spectra were extracted both using large 
apertures centered on the nebula and using a series of narrow apertures 
spanning the full spatial extent of the diffuse emission.  
We applied a wavelength calibration determined using HeNeArCdZn comparison lamp exposures and checked the accuracy of the 
solution using a number of sky lines.  Small linear shifts in the wavelength solution relative to the sky lines 
were measured and removed.  Details on the final spectral resolution, spectral range, and wavelength solution 
are listed in Table~\ref{tab:redux}.

Our subsequent analysis relies on having an accurate wavelength calibration, particularly in the regions around 
the \lya\ and \heii\ emission lines.  Therefore, as a further check on the accuracy of the wavelength solution, we cross-correlated 
our calibrated sky spectrum with an independently calibrated sky spectrum from 
a different program but the same telescope and instrument set-up.  
The two independently calibrated datasets 
show no relative shifts between the \lya\ and \heii\ regions of the spectrum. 
 
Flux calibration was applied using standard star exposures of Feige~34 and Wolf~1346 for the 
2009 dataset, and Feige~110 and BD+33~2642 for the 2010 
dataset.\footnote{KPNO IRS Standard Star Manual; \citet{massey90}.}  
The sensitivity functions derived independently for the two datasets 
agree to within 17\% at the wavelength of \lya\ 
(at $\approx$3250\AA, near the atmospheric cut-off) and to 
within 1-7\% at the wavelengths of \civ, \heii, and \ciii.  
The larger discrepancy at the location of \lya\ is likely due to 
the very blue observed wavelength at this redshift ($\lambda_{obs}\approx3250$\AA), 
which means this region of the spectrum is more affected by atmospheric absorption, slit losses, 
and atmospheric dispersion, as well as larger flat-fielding errors due to the typically red 
color of flat-field lamps.

\section{Results}
\label{sec:results}

In this section we present the 2D spectra of PRG1 followed by the 
surface brightness, emission line, and kinematic profiles.  
Throughout our discussion, a positive velocity offset refers to a redshift of 
\lya\ relative to the centroid of the \heii$\lambda$1640 line. 

\subsection{Spectra}
\label{sec:spectra}

\begin{figure}[h]
\includegraphics[angle=0,width=7in]{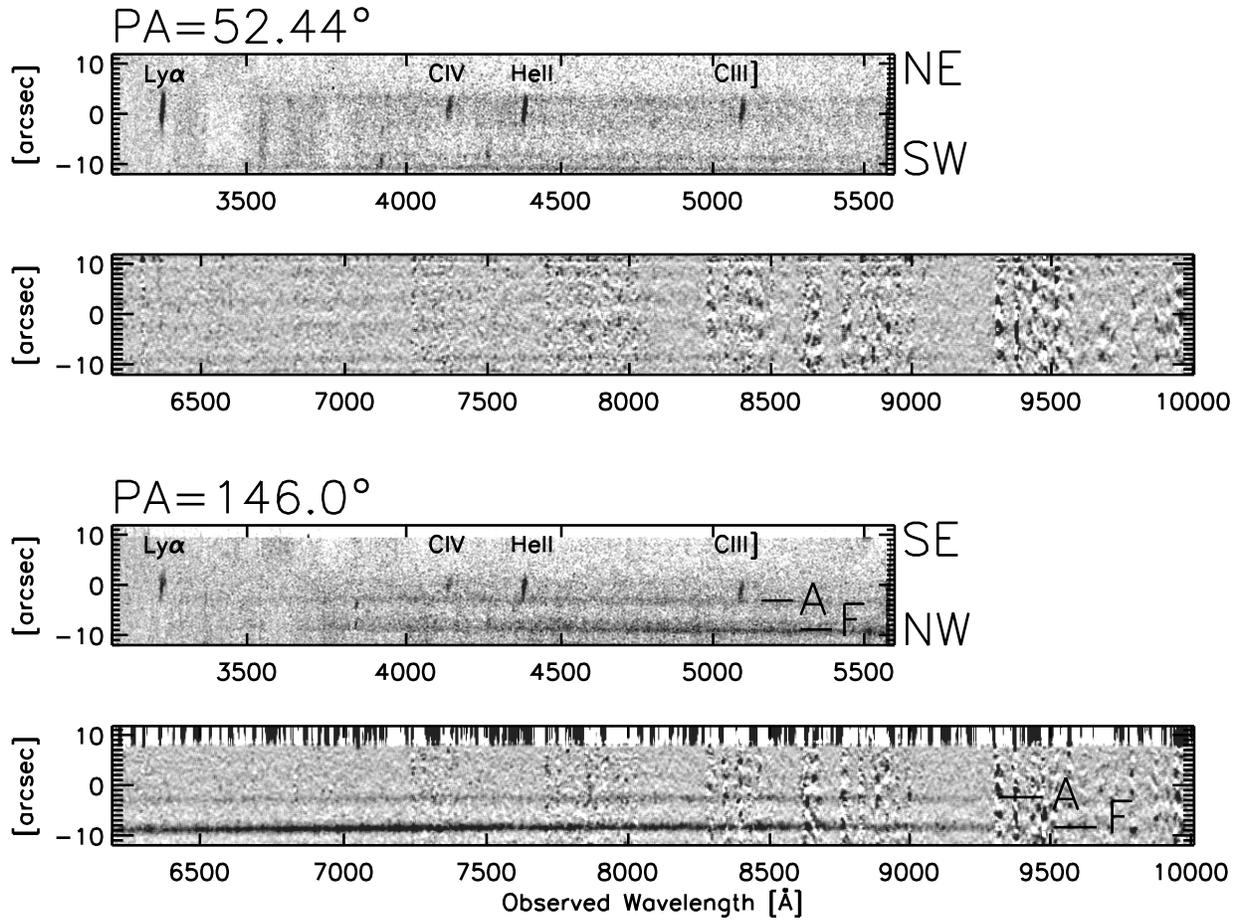} 
\caption[]{
Individual 2D spectra at PA=52\fdg44 and PA=146\fdg0 from the 2009 run 
(binned by two in the spectral dimension, to match the 2010 data) 
prior to flux calibration.  
Emission lines are labeled along with the positions of Sources `A' and `F' on the slit.  
Zero in the spatial direction corresponds to the position where 
the two slits cross, as listed in Table~\ref{tab:specobs}.
}
\label{fig:twodspec2009}
\end{figure}

\begin{figure}[h]
\includegraphics[angle=0,width=7in]{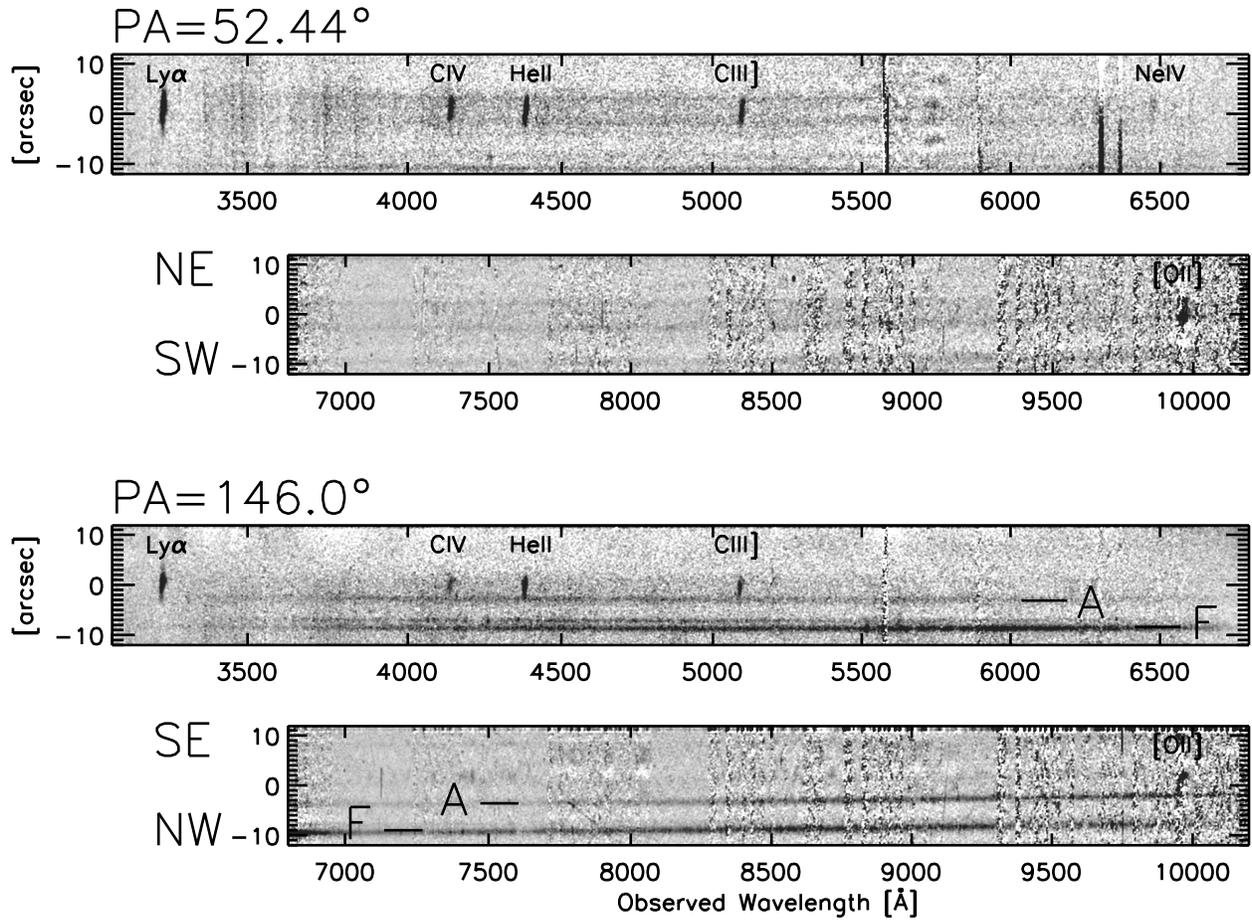} 
\caption[]{
Individual 2D spectra at PA=52\fdg44 and PA=146\fdg0 
from the 2010 run prior to flux calibration.  
Emission lines are labeled along with the positions of Sources `A' and `F' on the slit.  
Zero in the spatial direction corresponds to the position where the two slits cross, as 
listed in Table~\ref{tab:specobs}.  
}
\label{fig:twodspec2010}
\end{figure}

\begin{figure}
\center
\includegraphics[angle=0,width=6.5in]{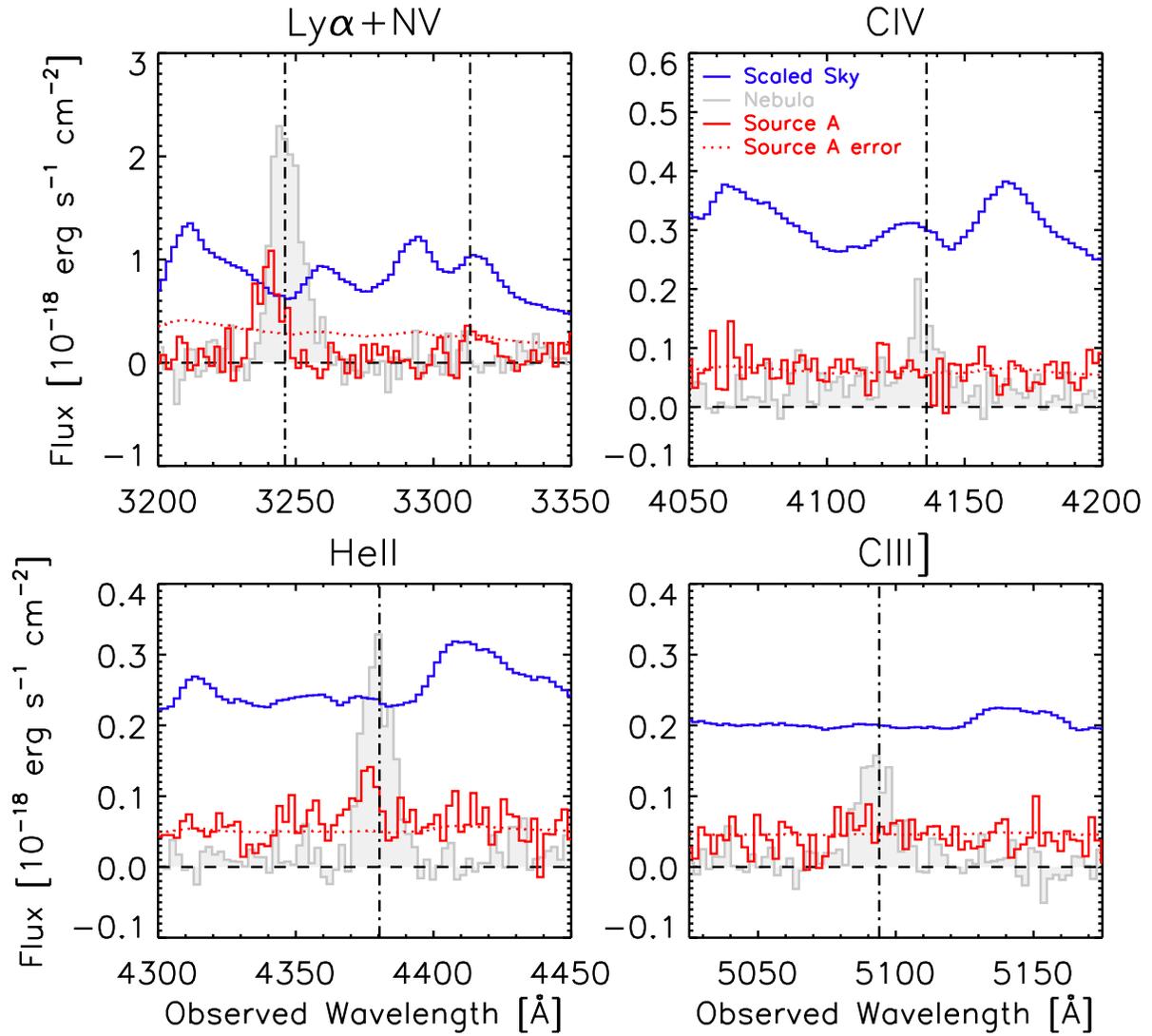} 
\caption[]{
Source A spectrum at the locations of \lya, \nv, \civ, \heii, and \ciii\ at the redshift of PRG1, extracted from a 5 pixel (0.67\arcsec) aperture (red solid line), with the corresponding error spectrum overplotted (red dotted line).  The spectrum taken from the center of the nebula is shown as the filled grey region, and a scaled sky spectrum is shown in blue.  The spectrum of Source A shows strong \lya\ emission and a tentative detection of \heii, most likely emission from the nebula overlapping the position of Source A.  
There is no independent evidence from the spectrum of Source A that confirms that it lies at the redshift of PRG1. However, based on its proximity and the lack of continuum emission at $\lambda_{\rm obs}<3230$\AA, it is plausible that this source is associated with PRG1.
}
\label{fig:sourceAz}
\end{figure}

Figures~\ref{fig:twodspec2009}-\ref{fig:twodspec2010} show the 2D spectra for both the 2009 
and 2010 observing runs prior to flux calibration.  The \lya$\lambda$1216, 
\civ$\lambda$1550, \heii$\lambda$1640, and \ciii$\lambda$1909 
(as well as \oii$\lambda$3727 in the 2010 data) are clearly detected in both position angles.  
Continuum emission is detected from the diffuse nebula as well 
as from several nearby compact sources.  
The brightest galaxy intersected by the PA=146\fdg0 slit (labeled `F') 
is a foreground source at $z\approx0.479$ object, with unambiguous 
\oii, \hbeta, and \oiii\ emission visible in the LRIS spectroscopy.
The second brightest continuum source detected in the PA=146\fdg0 data, labeled `A' 
in Figures~\ref{fig:twodspec2009}-\ref{fig:twodspec2010}, is located at the northwest edge of the 
diffuse line emission.  The 1D spectral extraction at the position of Source A (Figure~\ref{fig:sourceAz}) shows clear \lya\ emission 
and possibly faint \heii\ emission, likely coming from the nebula rather than from the galaxy itself, 
but no other strong lines.  
While there appears to be a hint of emission in the Source A spectrum at the position of \nv\ at the nebula redshift, 
the lack of any corresponding emission at \civ\ or \ciii\ and the presence of a sky line at exactly the same spectral location leads us 
to conclude that this is not a real detection of \nv.  
We therefore have no strong spectral constraints on the redshift of Source A, but due to its proximity to 
the nebula and the lack of continuum emission shortward of \lya, it seems likely that Source A 
is a galaxy associated with PRG1.  

In what follows we combine the two years of data.  
The derived sensitivity functions between 
the two runs agree reasonably well, and the \heii\ flux measurements observed from PRG1 generally 
agree to within the errors for both position angles.  Therefore, we combine the data employing a simple 
variance-weighted mean after binning the 2009 data by two spectrally to match the 2010 data.  
We focus in this paper on the restframe UV lines; the \oii\ emission will be analyzed 
in the companion paper on the energetics and physical conditions within PRG1.

\subsection{Surface Brightness Profiles}
\label{sec:spatialprofile}

\begin{figure}[h]
\includegraphics[angle=0,width=6in]{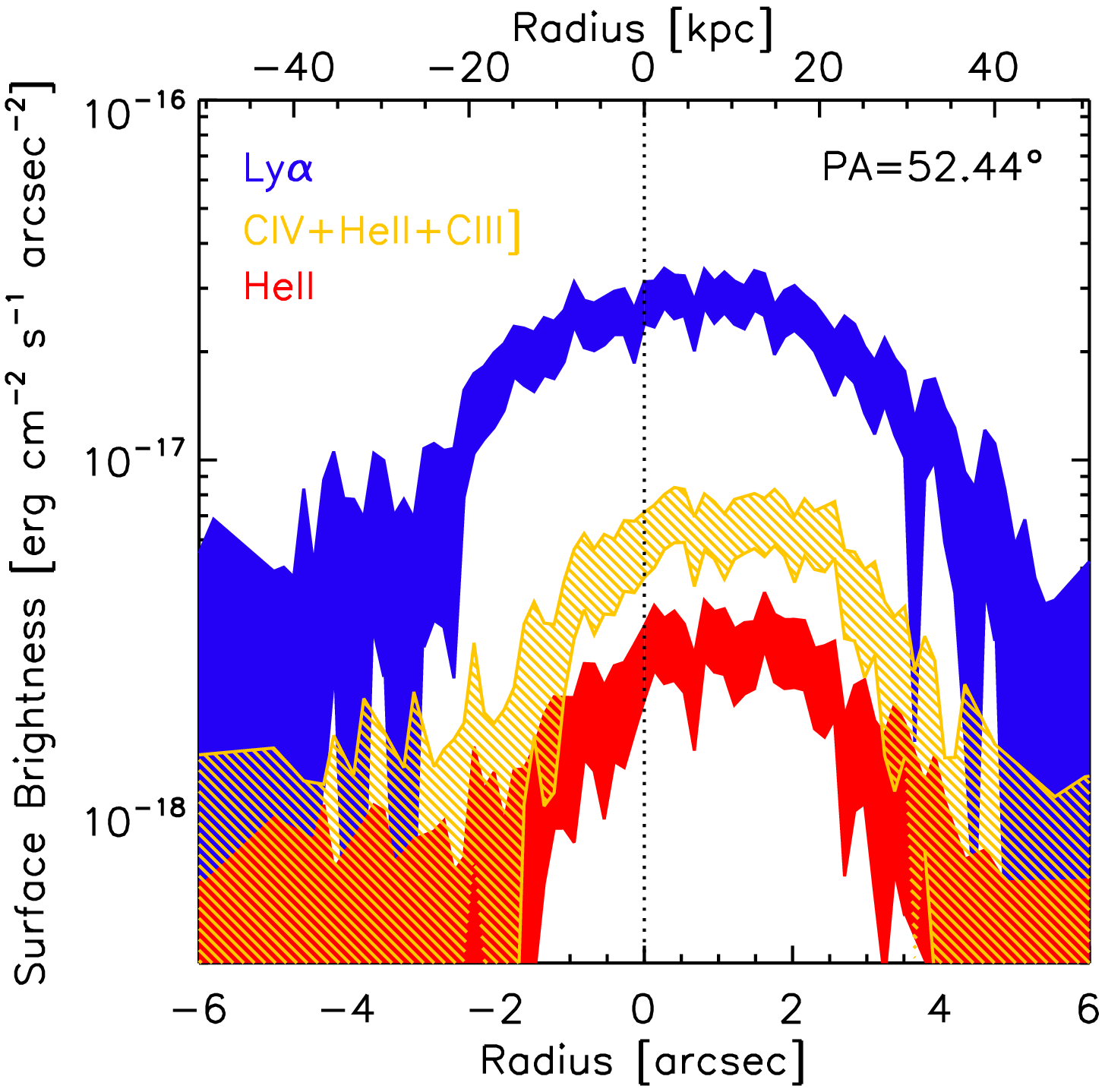} 
\caption[]{
Surface brightness profiles of \lya\ (solid blue region), \heii\ (solid red region), 
and a combined \civ+\heii+\ciii\ (hashed yellow region) along the PA=52\fdg44 slit.  
The filled regions span the range between the upper and lower error bars for each bin.  
The plot is centered in the spatial direction about the position where the two slits cross (dotted line).  For clarity, we 
restrict the x-axis of the plot to the range over which we have good signal-to-noise ratio measurements. 
}
\label{fig:spatialMID2010}
\end{figure}

\begin{figure}[h]
\includegraphics[angle=0,width=6in]{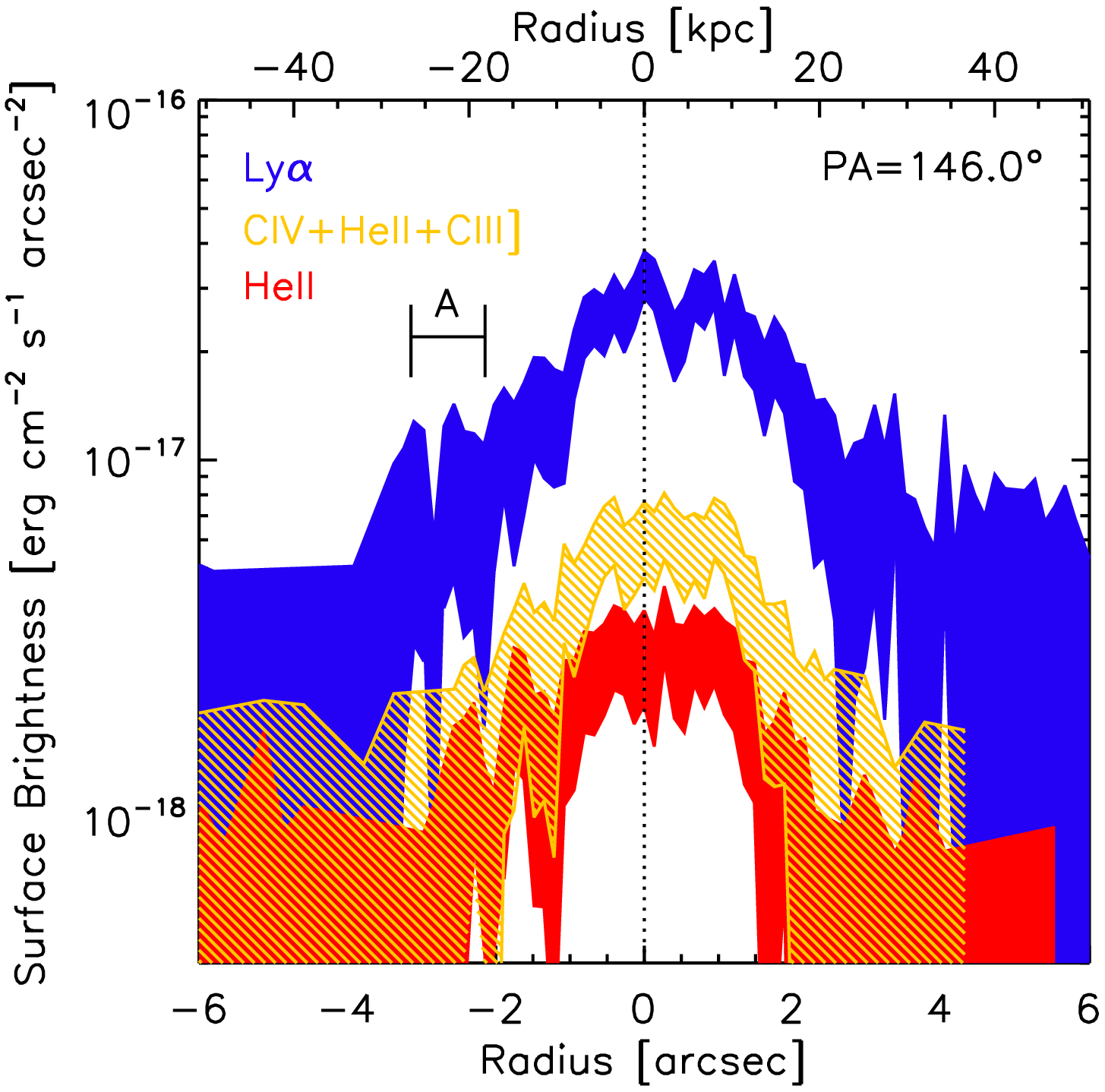} 
\caption[]{
Surface brightness profiles of \lya\ (solid blue region), \heii\ (solid red region), and a combined \civ+\heii+\ciii\ (hashed yellow region)  
along the 
PA=146\fdg0 slit.  
The filled regions span the range between the upper and lower error bars for each bin.  
The plot is centered in the spatial direction about the position where the two slits cross (dotted line).  For clarity, we 
restrict the x-axis of the plot to the range over which we have good signal-to-noise ratio measurements.  
The location of Source A is indicated.
}
\label{fig:spatialA2010}
\end{figure}

Surface brightness profiles are shown for both \lya\ and \heii\ as well as
for a stack of the \civ, \heii, and \ciii\ lines  
in Figures~\ref{fig:spatialMID2010}-\ref{fig:spatialA2010}.  
Measured above a surface brightness of $SB_{Ly\alpha}\approx4.5\times10^{-18}$ 
erg s$^{-1}$ cm$^{-2}$ arcsec$^{-2}$, 
the nebula spans almost 9.5\arcsec$\approx$80~kpc in diameter along the PA=52\fdg44 slit.  
To compare the relative size of the nebula in each tracer, we measure the diameters 
containing 50\% and 90\% of the total line luminosity (Table~\ref{tab:sbmodel}).  
The surface brightness profiles in \lya, the non-resonant \heii\ line, and 
the \civ+\heii+\ciii\ composite are all strikingly similar; 
for example in terms of half-light diameter, $D_{50}$, the \lya\ emission 
is only slightly more extended than the other rest-frame UV lines, 
by a factor of $\sim1.3$.  This result is consistent with what we found previously using 
shallower data \citep{pres09}.

\subsection{Emission Line Profiles}
\label{sec:lineprofile}

\begin{figure}[h]
\includegraphics[angle=0,width=5.5in]{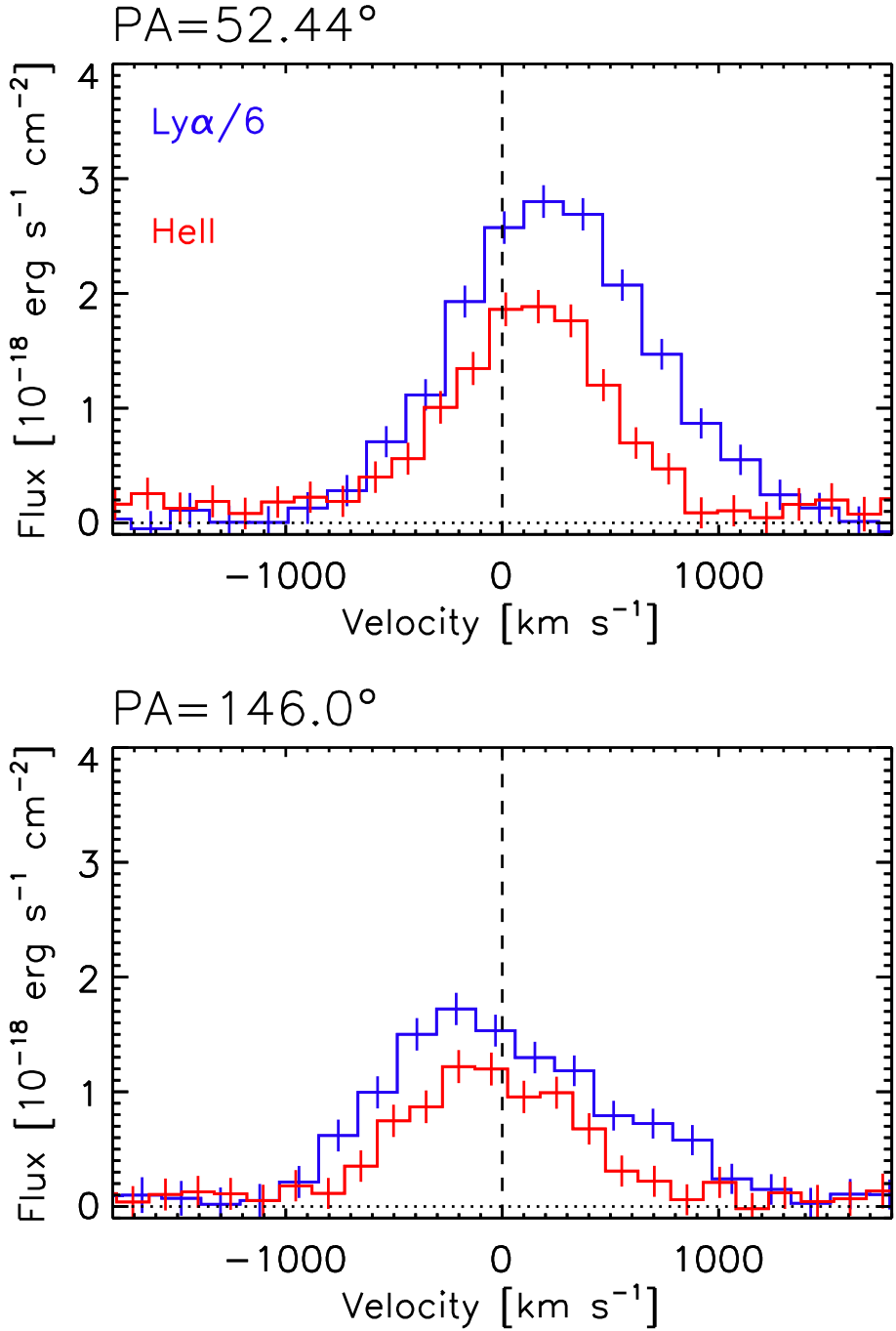} 
\caption[]{
Emission line profiles of \lya\ (blue line, scaled down for clarity) and \heii\ (red line) 
measured within \apsizeMID\arcsec\ and \apsizeA\arcsec\ wide apertures, 
respectively, at PA=52\fdg44 and PA=146\fdg0.
The vertical dashed line in each panel corresponds to the 
systemic velocity defined as the centroid of \heii\ 
at the position on the nebula where the two slits cross.
A positive velocity corresponds to a redshift relative 
to the systemic velocity.  
}
\label{fig:spectral2010}
\end{figure}

\begin{figure}
\includegraphics[angle=0,width=5.5in]{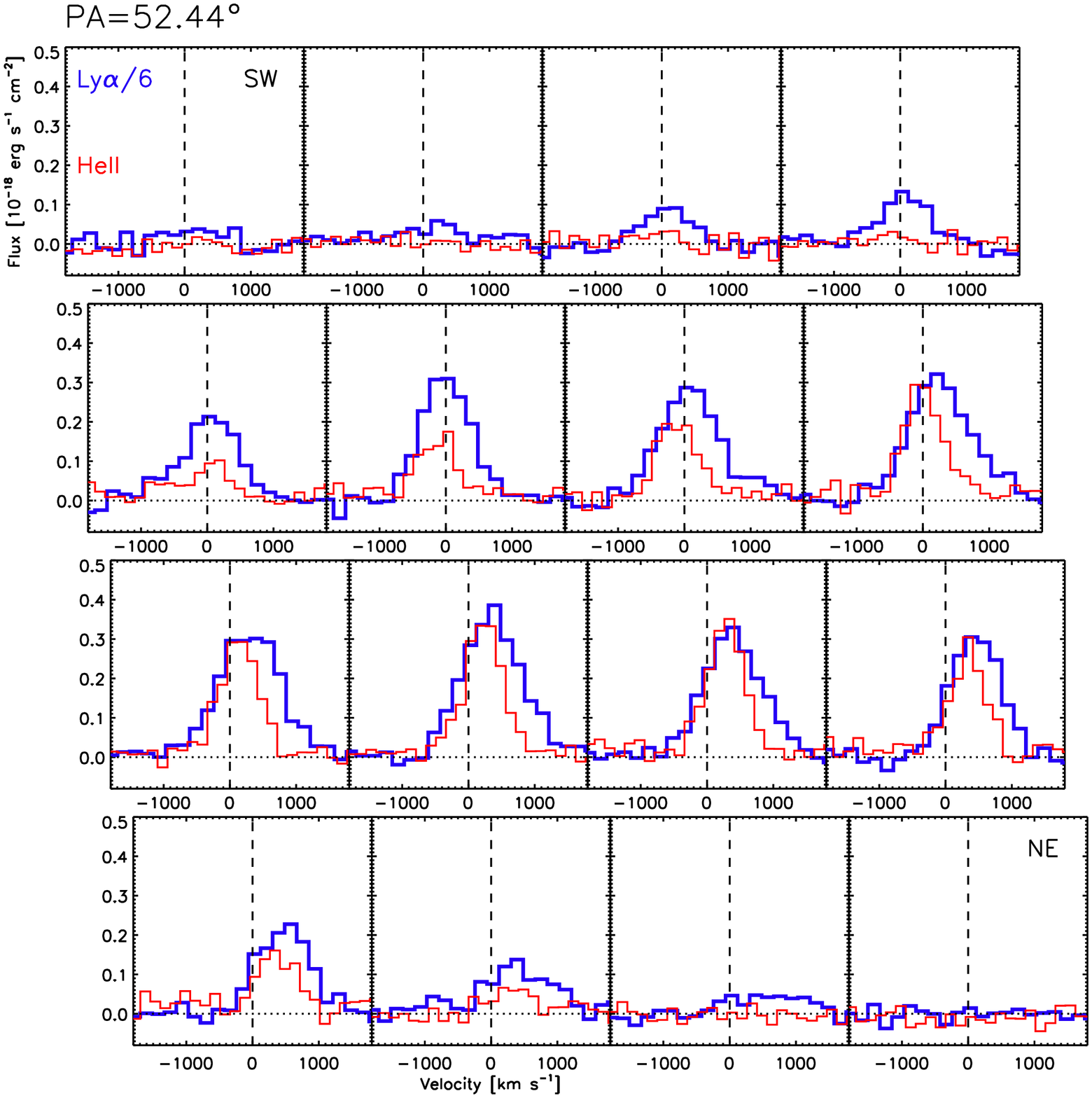} 
\caption[]{
Emission line profiles of \lya\ (blue line, scaled down for clarity) and \heii\ (red line) 
measured as a function of position from the SW (upper left) to the NE (lower right) end of the  
PA=52\fdg44 slit, extracted in 5~pix$=$0.67\arcsec\ apertures.  
The vertical dashed line in each panel corresponds to the 
systemic velocity defined as the centroid of \heii\ 
at the position on the nebula where the two slits cross.
A positive velocity corresponds to a redshift relative to the systemic velocity.   
}
\label{fig:spectralsliceMID2010}
\end{figure}

\begin{figure}
\includegraphics[angle=0,width=5.5in]{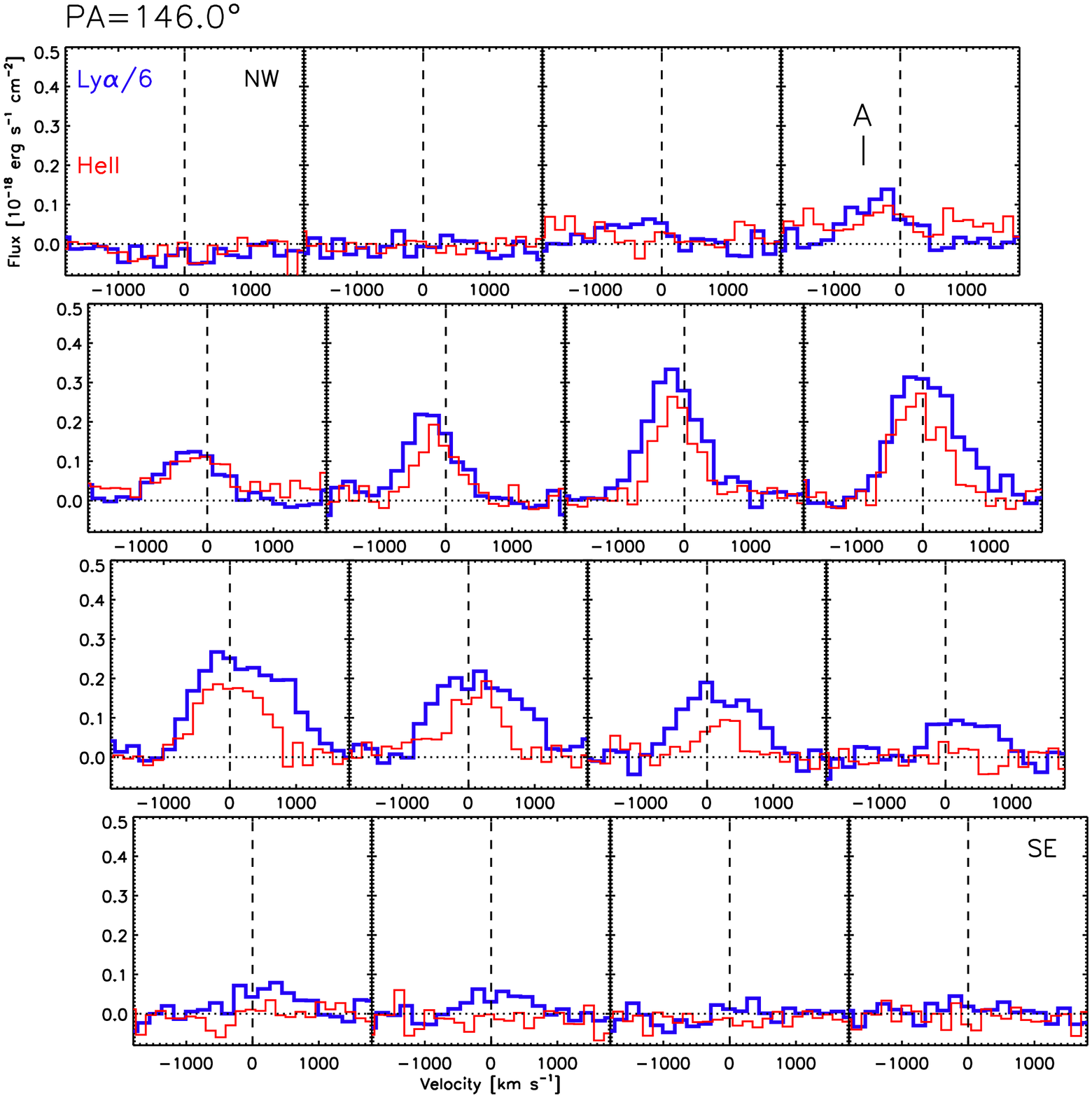} 
\caption[]{
Emission line profiles of \lya\ (blue line, scaled down for clarity) and \heii\ (red line) measured as a function of position 
from the NW (upper left) to the SE (lower right) end of the 
PA=146\fdg0 
slit, extracted in 5~pix$=$0.67\arcsec\ apertures.  
The vertical dashed line in each panel corresponds to the 
systemic velocity defined as the centroid of \heii\ 
at the position on the nebula where the two slits cross.
A positive velocity corresponds to a redshift relative to the systemic velocity.  
The position of Source A is indicated. 
}
\label{fig:spectralsliceA2010}
\end{figure}

Figure~\ref{fig:spectral2010} shows a comparison between the emission line 
profiles of \lya\ and \heii\ derived from wide extraction apertures chosen to maximize 
the total signal-to-noise ratio of all four strong lines 
(\apsizeMID\arcsec=\apsizeMIDkpc~kpc and \apsizeA\arcsec=\apsizeAkpc~kpc, 
respectively, for the PA=52\fdg44 and PA=146\fdg0 data).  
Figures~\ref{fig:spectralsliceMID2010}-\ref{fig:spectralsliceA2010} are multi-panel figures 
showing the same emission line profile comparison but as a function of position along the slit, 
where the individual panels corresponds to subapertures (5~pix$\approx0.67$\arcsec) spanning the full extent of the nebula. 
The vertical dashed lines represent the systemic velocity, defined as the centroid 
of the \heii\ line at the position where the two slits cross.  
Our spectral resolution is not sufficient to resolve multiple peaks in the \lya\ emission line due 
to radiative transfer effects, 
so instead we focus on the centroid offsets of the \lya\ line relative to the non-resonant \heii\ line.  

\begin{figure}[h]
\includegraphics[angle=0,width=5.5in]{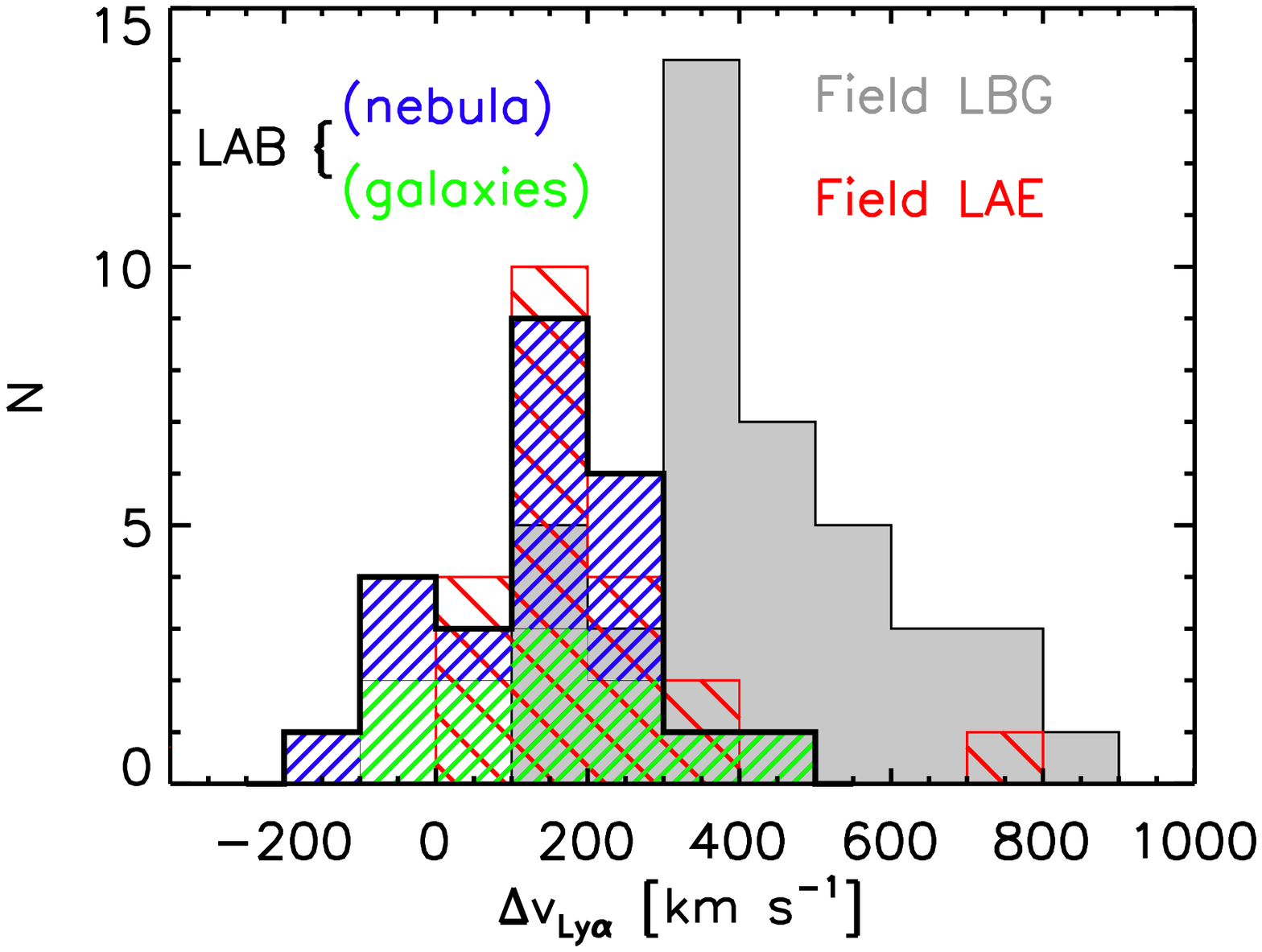} 
\caption[]{
Velocity offset between \lya\ and the systemic velocity for 
LBGs \citep[filled grey histogram;][]{stei10} and LAEs 
\citep[narrow hashed red histogram;][]{mcl11,guaita13,hashimoto13,song14}.  
Velocity offsets measured within LABs are shown, as measured at the position of 
embedded galaxies \citep[slanted hashed green histogram;][]{francis96,mcl13,yang14b} and from spatial apertures within the extended nebula \citep[slanted hashed blue histogram; thirteen measurements are from this work on PRG1, measured within 5~pix$\approx0.67\arcsec$ spatial apertures with a minimum of $SNR=3$ in both \lya\ and \heii, and one additional measurement is taken from LABd05 by][]{yang14a}.  \lya\ nebulae show \lya\ velocity offsets that are lower than those measured for the bulk of the LBG population but similar or slightly lower than what is seen in LAEs.
}
\label{fig:velooffset}
\end{figure}

In Figure~\ref{fig:velooffset}, we plot the velocity offset of \lya\ measured in 
thirteen spatial apertures within the nebula 
(5~pix$\approx$0.67\arcsec, with a minimum $SNR=3$ in both \lya\ and \heii) 
and compare with similar measurements from other galaxy populations.  
The typical observed offset between the two lines is 
$\Delta v \equiv v_{\rm Ly\alpha}-v_{\rm HeII} \sim 100-200~{\rm km~s^{-1}}$, 
depending on position.  
This low velocity offset is similar to what has been seen in other \lya\ nebulae \citep{yang11a,mcl13}, 
however, here we are able to probe the kinematics point-by-point within the extended gas to show that the 
velocity offset is consistently low across the entire $\sim$80~kpc nebula.  
The measured velocity offset is less than what is seen in 
LBGs \citep{stei10} and more similar to that observed in \lya-emitting galaxies 
\citep{mcl11,hashimoto13,guaita13,song14}.   

\subsection{Kinematic Profiles}
\label{sec:veloprofile}

\begin{figure}[h]
\includegraphics[angle=0,width=5.5in]{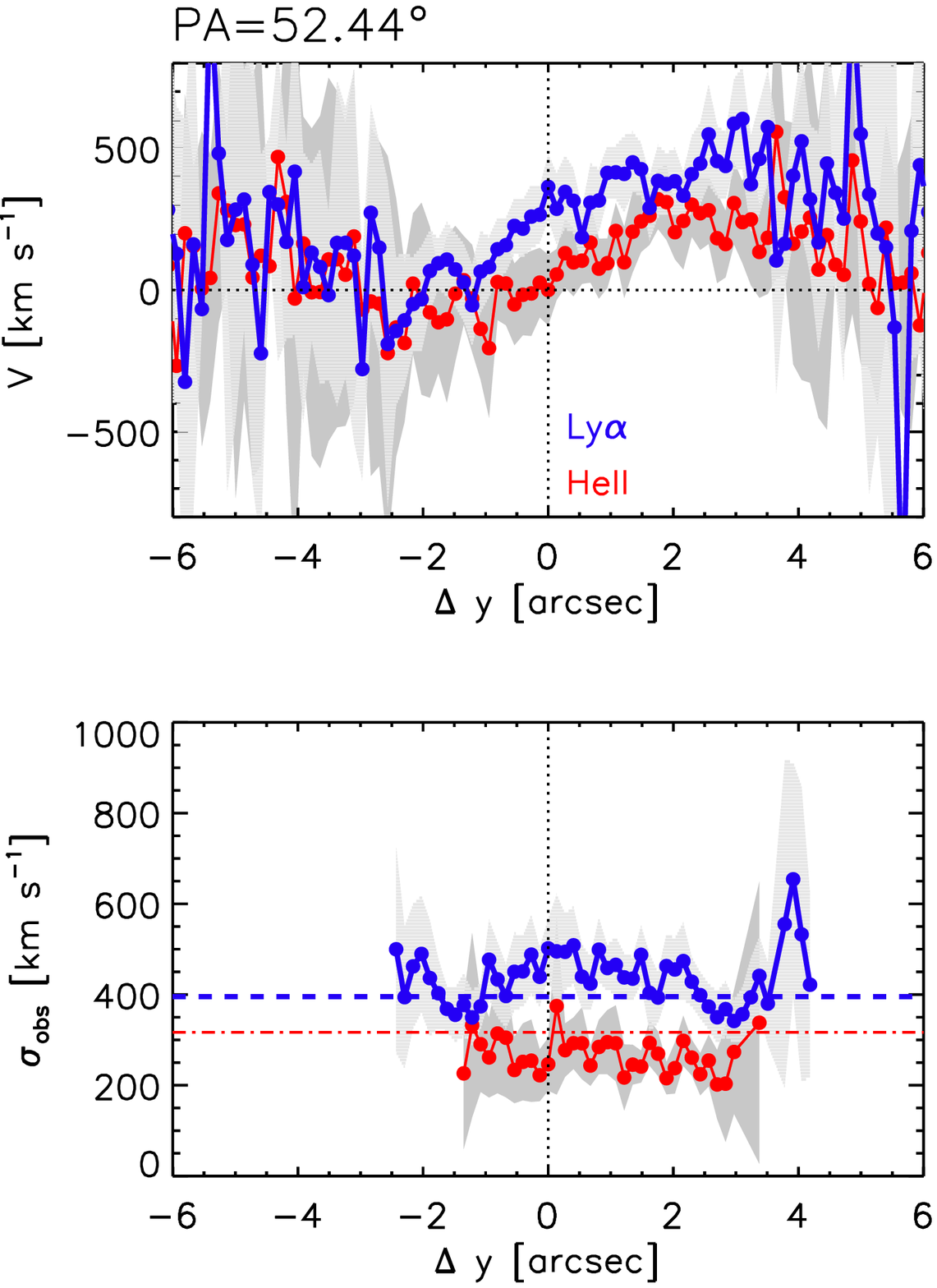} 
\caption[]{
Rotation (top panel) and velocity dispersion (bottom panel) curves for \lya\ and \heii\ for the  PA=52\fdg44.  
\lya\ is shown as a thick blue line, and \heii\ is shown as a thin red line. 
The hashed grey and grey-shaded regions indicate the corresponding error bars.  
The velocity dispersion is the observed value, with the instrumental resolution at the 
location of \lya\ and \heii\ shown as blue dashed and red dot-dashed lines, respectively. 
}
\label{fig:rotcurvesMID2010}
\end{figure}

\begin{figure}[h]
\includegraphics[angle=0,width=5.5in]{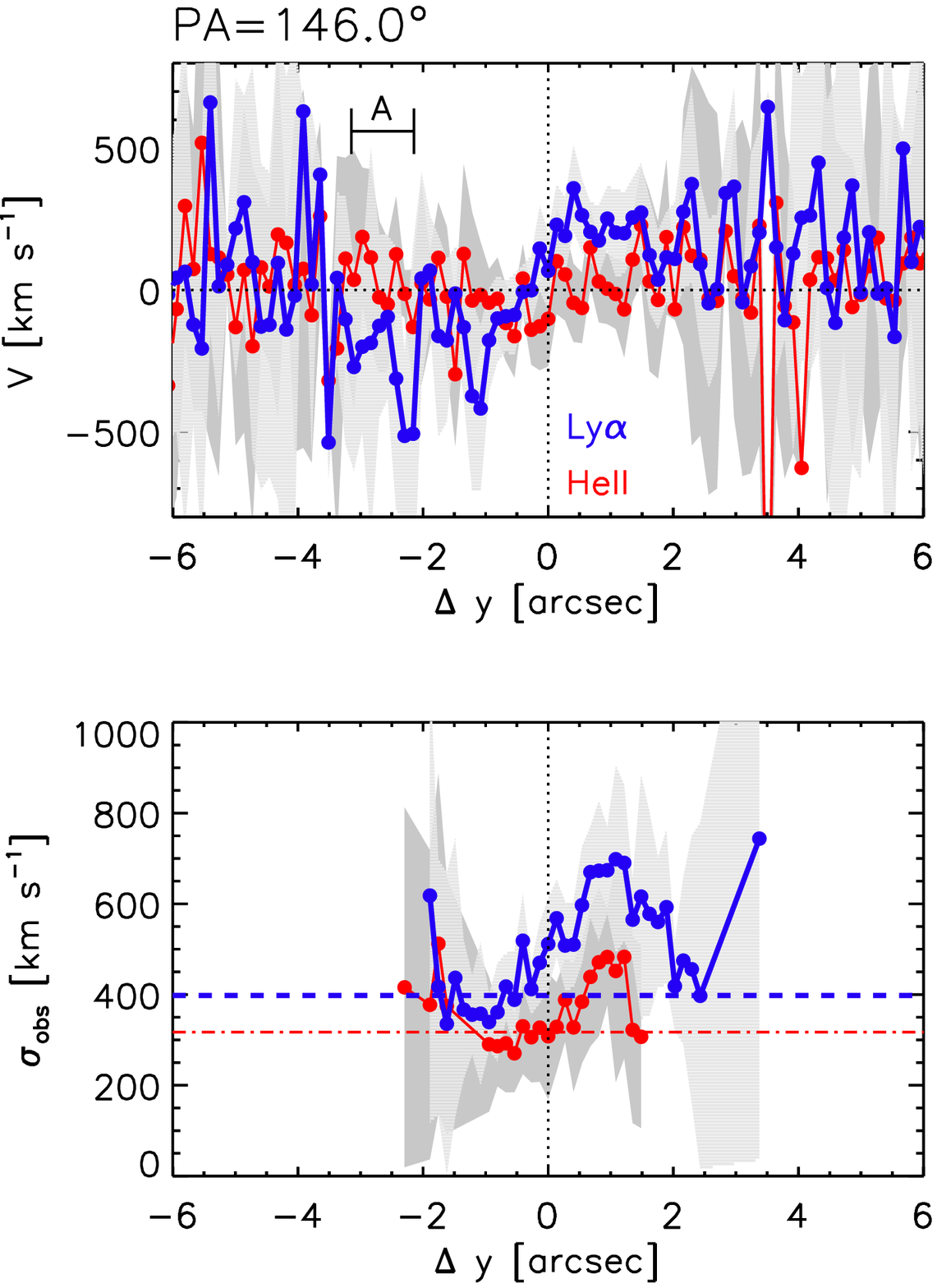} 
\caption[]{
Rotation (top panel) and velocity dispersion (bottom panel) curves for \lya\ and \heii\ for the  
PA=146\fdg0.  
\lya\ is shown as a thick blue line, and \heii\ is shown as a thin red line.  
The hashed grey and grey-shaded regions indicate the corresponding error bars.  
The velocity dispersion is the observed value, with the instrumental resolution at the 
location of \lya\ and \heii\ shown as blue dashed and red dot-dashed lines, respectively. 
}
\label{fig:rotcurvesA2010}
\end{figure}

In Figures~\ref{fig:rotcurvesMID2010}-\ref{fig:rotcurvesA2010}, 
we present the velocity and velocity dispersion profiles for \lya\ and \heii.  
Again, we see very small velocity offsets 
between \lya\ and \heii; 
typically in the range of $\sim100-200$ km~s$^{-1}$ to the red.  
At the same time, the large-scale velocity 
profile in the PA=52\fdg44 slit shows a coherent velocity gradient -- $\sim500$ km~s$^{-1}$ 
over the central 50 kpc of the nebula -- while in the PA=146\fdg0 slit, the velocity profile 
is much shallower.  
While in most spatial apertures the kinematic offset is to the red, in PA=146\fdg0 on the 
side of the nebula closest to Source A there is a hint of a reversal, i.e., \lya\ is slightly 
offset to the blue.  

The linewidths measured as a function of position are shown in the lower panels of 
Figures~\ref{fig:rotcurvesMID2010}-\ref{fig:rotcurvesA2010}.  
Across the PA=52\fdg44 slit, the linewidth profile is quite flat, only marginally resolved in \lya, 
and unresolved in \heii.  
On the other hand, in PA=146\fdg0 we see a peak in the linewidth profile that is clearly resolved and 
spatially coincident in both \lya\ and \heii.  In later sections, we use these linewidths as a measure 
of the velocity dispersion as a function of position within the nebula.  However, given the spatial resolution 
of our data, we note that these measurements inevitably includes a contribution from macroscopic 
velocity gradients, i.e., of order 80 km~s$^{-1}$ across the 1\arcsec\ seeing disk for PA=52\fdg44.

\section{Discussion}
\label{sec:discussion}

Using the results of the previous section, we can now address the questions posed in Section~\ref{sec:intro}: 
what is the role of resonant scattering in creating the extended \lya\ nebula, and what are the 
kinematics of the gas?  We then discuss the implications of these results for the physical model and 
underlying power source for PRG1.

\subsection{\lya\ Optical Depth: A Back-of-the-Envelope Estimate}
\label{sec:lyaoptdepth}

We expect that \lya\ will be optically thick under all but the most extreme situations, yet 
the consistently small velocity offsets we observe between \lya\ and the non-resonant \heii\ line 
suggest that \lya\ is not being substantially affected by complex radiative transfer in PRG1.  
This implies either that \lya\ is optically thin (which we will demonstrate in this section 
is not likely the case), or that \lya\ is produced in situ over an extended area, thereby largely 
avoiding the effects resonant scattering would impart.

Before discussing our observations in more detail, it is useful to review 
how optically thin \lya\ can be under plausible astrophysical conditions, particularly 
under the limiting case of a highly ionized medium.  

We start by making the assumption of ionization equilibrium:

\begin{eqnarray}
\alpha_{B} n_{e} n_{HII} = \Gamma n_{HI} \\
\Gamma \approx Q \sigma_{p} /(4\pi R^{2}) 
\end{eqnarray}

where $n_{e}$ is the electron number density (cm$^{-3}$), n$_{HI}$ is the HI number density (cm$^{-3}$), n$_{HII}$ is the HII number density (cm$^{-3}$),
$\Gamma$ is the photoionization rate (photoionizations/s), $Q$ is the luminosity of ionizing photons (photons/s),
$R$ is the radius from ionizing source (cm), $\sigma_{p}$ is the photoionization cross-section (cm$^{2}$),
and $\alpha_{B}$ is the Case B recombination coefficient for
$HI$.

The \lya\ optical depth is:

\begin{eqnarray}
\tau_{Lya} = n_{HI}  \sigma_{Lya}  L
\end{eqnarray}

where $\tau_{Lya}$ is the optical depth at line center, $n_{HI}$ is the HI number density (cm$^{-3}$),
$\sigma_{Lya}$ is the Lya absorption cross-section (cm$^{2}$), and $L$ is the path through system (cm).

We make the approximation that H is highly ionized (e.g., in the region around an AGN):
\begin{eqnarray}
n_{HII} \sim n_{H} \sim n_{e} 
\end{eqnarray}

Substituting into equation (1), we obtain the following relation:
\begin{eqnarray}
n_{HI} & \approx & \alpha_{B} n_{H}^2 / \Gamma \\
       & \approx & 4 \pi R^2 \alpha_{B} n_{H}^2 / (Q \sigma_{p})
\end{eqnarray}

We rewrite this in terms of the ionization parameter $U = Q/(4 \pi R^2 c n_{H})$, where $c$ is the speed of light: 
\begin{eqnarray}
n_{HI} \approx \alpha_{B} n_{H}/(c \sigma_{p} U)
\end{eqnarray}

Combining this relation with 
equation (3) yields: 

\begin{eqnarray}
\tau_{Lya} \approx \alpha_{B} n_{H} \sigma_{Lya} L /(c \sigma_{p} U)
\end{eqnarray}

We adopt the following values: 
$\alpha_{B} = 2.59\times10^{-13}$ cm$^{3}$ s$^{-1}$, 
$\sigma_{p} = 6.3\times10^{-18}$ cm$^{2}$ at 1 Ryd, 
$\sigma_{Lya} =  5.9\times10^{-14}$ cm$^{2}$ 
\citep{storey1995,verner96,ost89}, 
$c=3.00\times10^{10}$ cm s$^{-1}$, 
$L = 50$~kpc and $n_{H} = 1.0$ cm$^{-3}$.  

Taking a range of ionization parameter values ($U = 10^{-3}-1$) 
results in a \lya\ optical depth of: 
\begin{equation}
\tau_{Lya}\approx1.2\times10^{7}-1.2\times10^{4}
\end{equation}
in the limit of high ionization.  

In order for $\tau_{Lya} < 1$, we would therefore need one of the following to be true:
\begin{itemize}
\item $U \gtrsim 10^{4}$, which is many orders of magnitude higher than what is measured for a typical AGN broad line region \citep{peterson97}.  
\item $n_{H} \lesssim 10^{-7}-10^{-4}$ cm$^{-3}$, which is 
much less than the (albeit uncertain) existing measurements of $n_{H}$ for \lya\ nebulae 
\citep[$\sim$1-30 cm s$^{-3}$; e.g.,][]{dey05,pres09}.
\item $L \lesssim 5\times10^{-3}-5$ pc, i.e., the \lya\ is emerging from a very thin skin.  
This could arise if 
(a) the \lya\ we observe is produced via photoionization within a 
``blister HII region" illuminated by an offset source, or if 
(b) \lya\ is produced throughout the cloud but all buried \lya\ is efficiently 
extinguished, i.e., by dust.
\end{itemize}

Thus, even in the highly ionized limit, $\tau_{Lya} >> 1$ for most reasonable physical parameters.
Given sufficient spectral resolution, we would expect to see evidence for substantial resonant 
scattering of the \lya\ line in the form of a larger spatial extent and/or intrinsically double 
peaked, complex profiles with kinematic offsets relative to a non-resonant tracer.  
At lower spectral resolution, this would translate into broadened 
\lya\ emission lines with offsets in the observed line centroid.

\subsection{Extended Line Emission and the Role of Resonant Scattering}
\label{sec:extent}

How much is \lya\ being affected by radiative transfer effects relative 
to the other emission lines, or similarly, how much is \lya\ resonant scattering responsible 
for the large physical extent of the nebula seen in \lya?  
At the spectral resolution of our data, we would expect that any intrinsically complex, 
multi-peak \lya\ profiles to result in the observed \lya\ line being broader than a non-resonant 
tracer.  This is consistent with our observation that the \lya\ line is broader than \heii\ in the 
aperture where both lines are resolved 
($\sigma_{HeII}\sim390$ km~s$^{-1}$ and $\sigma_{Ly\alpha}\sim570$ km~s$^{-1}$ for \lya, 
after correcting for the instrumental resolution; 
Figure~\ref{fig:rotcurvesA2010}).  
This suggests that, as expected, \lya\ is more optically thick and undergoing more complicated 
radiative transfer than the \heii\ line.  

However, the simple observational fact that a non-resonant line like \heii\ 
is seen to be nearly as spatially extended as the \lya\ emission is a strong argument that resonant 
scattering of centrally-produced \lya\ is not the primary factor responsible for the large 
spatial extent of the \lya\ emission.  
A similar observation was made in the case of LABd05, a \lya\ nebula at $z\approx2.7$ that 
shows diffuse UV continuum emission comparable in extent to the \lya\ \citep{pres12b}.  
In at least these two systems, \lya\ scattering is not the main reason we observe 
a $\sim100$~kpc scale \lya\ nebula.  Instead, the \lya\ photons are predominantly being produced 
in situ within the extended gas.
At the same time, polarization data from a different \lya\ nebula system suggests a 
significant contribution from scattered Lya emission \citep[SSA22-LAB1;][]{hayes11}.  
The prevalence of scattering versus in situ production in \lya\ nebulae as a class remains to 
be quantified, but these few case studies suggest that both mechanisms play a role in producing 
extended \lya\ sources.
 
\subsection{Kinematics within the Spatially Extended Gas}
\label{sec:kinematics}

\lya\ radiative transfer modeling indicates that \lya\ photons propagating through 
outflowing or infalling gas should appear redshifted or blueshifted, respectively, 
relative to the systemic velocity \citep[e.g.,][]{dijk06a,ver06}.   
With this in mind, we can ask whether \lya\ shows velocity offsets relative to the non-resonant 
\heii\ line in PRG1.  
In general, we see relatively small offsets between 
the two lines (typically $\sim100-200$ km~s$^{-1}$), with \lya\ usually shifted to the red (suggestive of an outflow).  
In the case of a simple outflowing shell model, these offsets would correspond to expansion velocities of 
$\sim50-100$ km~s$^{-1}$ \citep[e.g.,][]{ver06}.  
In addition, a few spatial apertures (located near Source A) 
show blue-shifts of Ly$\alpha$ suggestive of mild infall.  
As we showed in Section~\ref{sec:lineprofile}, 
these velocity offsets are overall lower than what is seen in typical UV-selected star-forming galaxies, 
and comparable or perhaps slightly lower than for \lya-emitting galaxies.

By itself, the low velocity offsets could imply one of several possibilities.  
The outflow velocities could be intrinsically lower in these systems, or the column 
density of neutral gas could be low (either globally or due to local patchiness).  
Alternatively, since the radiative transfer from an extended source of emissivity 
generically results in observed kinematic offsets that are suppressed relative 
to the case of a central source \citep[e.g.,][]{ver06}, 
the low velocity offsets could simply reflect the fact that the \lya\ photons in the \lya\ 
nebula are being produced over an extended region, rather than being scattered from a central source.  
In this scenario, the \lya\ profiles in each aperture do not actually encode information 
about the full velocity structure of the system, but instead reflect only small local velocity 
offsets between the point of emission of the \lya\ photons and the final scattering location.  
For this reason, the \lya\ kinematics closely resemble what is measured using a non-resonant line generated within the same region.   
Combined with the results of the previous section, it seems clear that both the spatial structure 
and kinematics of the system are consistent with in situ production of \lya\ photons in PRG1.  
From a purely observational perspective, the fact that \lya\ traces the non-resonant \heii\ line 
so well, suggests that using \lya\ alone to do kinematic studies 
may actually be more reliable in this case than is often assumed.  

\subsection{Towards a Physical Model for PRG1: Evidence for Large-scale Rotation}
\label{sec:giantdisk}
 
The deep spectroscopy allows us to probe the kinematics of the diffuse gas using multiple lines with great sensitivity out to large 
physical scales.  
A successful physical model for the gas kinematics in PRG1 must be 
able to explain the following observations: (1) a pronounced monotonic velocity gradient in the PA=52\fdg44 
data, with a flattening at large radii, and a relatively flat velocity profile in PA=146\fdg0; 
(2) a consistently low velocity dispersion in PA=52\fdg44 
($\sigma_{Ly\alpha}\lesssim300$ km~s$^{-1}$, corrected for the instrumental resolution), 
and (3) a conspicuous resolved peak in the velocity dispersion profiles for both \heii\ and \lya\ in 
the PA=146\fdg0 slit, corresponding to $\sigma_{HeII}\sim390$ km~s$^{-1}$ and $\sigma_{Ly\alpha}\sim570$ km~s$^{-1}$.

\begin{figure}[h]
\includegraphics[angle=0,width=5in]{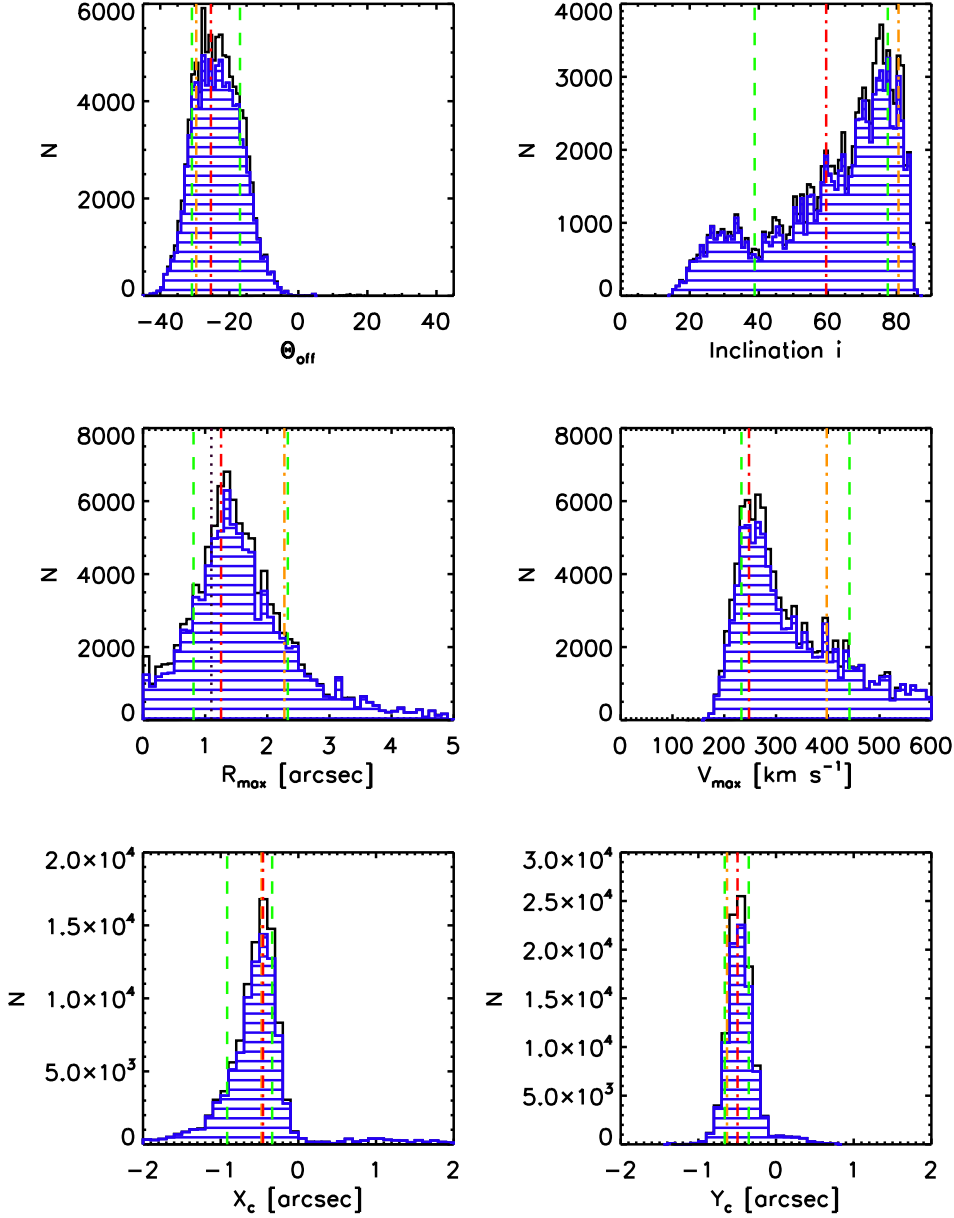} 
\caption[]{
Posterior distributions derived using a Markov Chain Monte Carlo fitting approach 
and a simple thin disk model with six parameters: $\Theta_{off}$, the angle between 
the PA=52.44\degree\ slit and the major axis of the disk (top left), $i$, the inclination 
of the disk relative to face-on (top right), $R_{max}$, the radius at which the disk 
flattens to the maximum velocity (middle left), $V_{max}$, the maximum velocity of the 
disk (middle right), and $X_{c}$ and $Y{c}$, the offsets of the slit crossover point 
relative to the disk center.  
The green dashed lines indicate the 67\% confidence intervals quoted 
in Table~\ref{tab:toymodel}, and the orange and red dot-dashed lines show the two random 
draws from the posterior distributions that are overplotted on the data in Figure~\ref{fig:toymodelbest}. 
The black dotted line in the $R_{max}$ panel corresponds to the maximum seeing during 
our spectroscopic observations.
}
\label{fig:posteriors}
\end{figure}

It is easiest to understand the velocity profile within the system if the nebula represents gas undergoing rotation.  
To show this, we construct a toy model of a simple thin disk with six parameters: 
the offset angle between the PA=52\fdg44 slit and the major axis of the disk ($\Theta_{off}$, 
between -45$^{\circ}$ to 45$^{\circ}$), the disk inclination ($i$, between 0$^{\circ}$ and 90$^{\circ}$), 
the maximum velocity of the disk ($V_{max}$, between 0 and 600 km~s$^{-1}$), 
the radius at which the disk reaches $V_{max}$ ($R_{max}$, between 0 and 6\arcsec), 
and the position of the slit crossover relative to the disk center 
($X_{c}$, $Y_{c}$, within $\pm2$\arcsec, 
where a positive offset in both parameters indicates a slit crossover located to the 
northwest of the disk center).  
We use a simple Markov Chain Monte Carlo (MCMC) fitting approach and 100,000 iterations 
to determine the best fit to the velocity profile in both slits, 
and estimate the posterior distribution for each parameter (Figure~\ref{fig:posteriors}).  
While we have shown that \lya\ and \heii\ show very similar behavior overall, there are small differences, particular 
in the region around Source A.  For the purposes of fitting the velocity profiles, therefore, 
we use the more reliable \heii\ line and restrict the fitting to only those apertures 
within the central $R<3.5\arcsec$ of the nebula.

\begin{figure}[h]
\includegraphics[angle=0,width=5in]{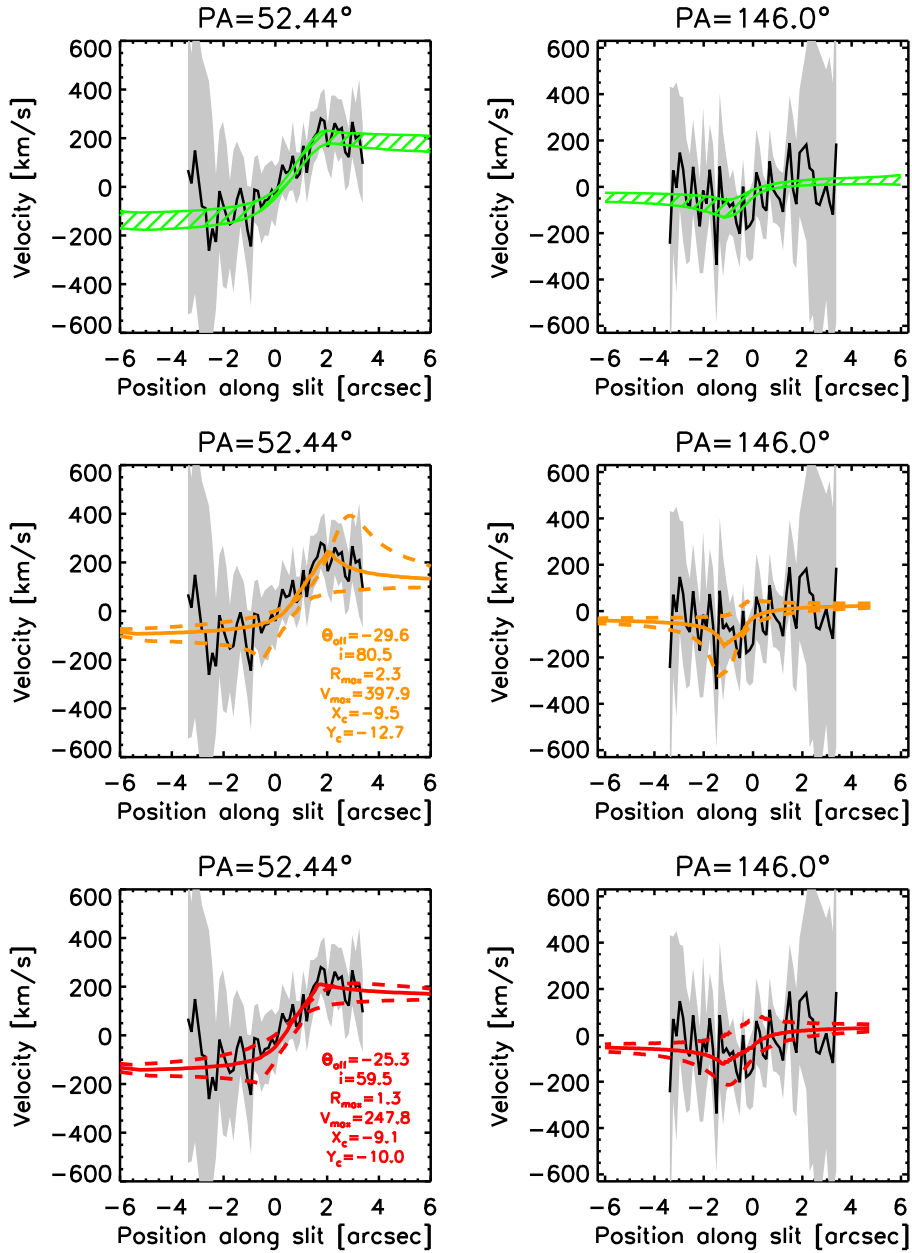} 
\caption[]{
The observed \heii\ velocity profiles in the PA=52\fdg44 (left) and 
PA=146\fdg0 (right) slits are shown with a solid black line and grey shade regions 
indicating the error bars.  Overplotted in color are velocity profile predictions from the 
thin disk toy model.  The green hashed bands in the top row represent the range of models 
spanned by the 67\% confidence intervals quoted in Table~\ref{tab:toymodel}.  
The lower two rows show two random draws from the posterior distributions 
(orange and red colored lines), as indicated using the same color coding in Figure~\ref{fig:posteriors}.  
The solid colored lines represent the velocity profile predicted along the center of 
the slit, while the dashed colored lines track the two slit edges.  
The corresponding model parameter values for each random draw are given in the left panel legend.  
}
\label{fig:toymodelbest}
\end{figure}

The posterior distributions suggest that the velocity profiles agree reasonably well with a thin disk model.  
In Table~\ref{tab:toymodel}, we list the 67\% confidence intervals for each parameter, and in 
Figure~\ref{fig:toymodelbest} we show the range of predicted velocity profiles corresponding to 
these confidence intervals as well as the predicted profiles for two random draws from 
the posterior distributions, overplotted on the data.  
The corresponding disk dynamical masses are of the order of M($<$R=20kpc)$\sim3\times10^{11}$ M$_{\odot}$. 
The fact that the major axis of the model disk is roughly aligned with PA=52\fdg44 explains 
the classic ``rotation curve'' structure seen in the velocity profile, i.e., the monotonic 
velocity gradient and 
the flattening at large radii.  Similarly, the PA=146\fdg0 slit lies roughly along the minor axis, explaining 
the flatter velocity profile.  We note that the surface brightness profiles of PRG1 
can be reasonably well fit with an exponential profile, consistent with this picture.

The velocity dispersion for a simple rotating disk observed at some inclination angle 
is a combination of the intrinsic velocity dispersion (e.g., due to turbulence) as well as 
the smearing of the disk rotation within the slit; 
the latter component typically leads to an observed peak in the velocity dispersion at the disk center.  
Our simple thin disk model does not allow for a formal prediction of the velocity dispersion, but 
the spread between the edges of the slit can be taken as an indication of the velocity dispersion 
that would be measured simply due to rotation smearing.  
For reasonable models, we find that the spread in velocity sampled by the slit is only of order 
200 km~s$^{-1}$ ($\sigma\lesssim100$ km~s$^{-1}$), well below the resolution of our data.  Thus, if 
the kinematics were driven solely by rotation in a thin disk, we would not expect to resolve the lines 
anywhere across the nebula.  
In addition, the MCMC fitting analysis strongly prefers a center of rotation that is offset with 
respect to the velocity dispersion peak by about 1-2\arcsec\ to the northwest.  
Together, these facts suggest the velocity dispersion peak we observe is not 
the kinematic center of the system, but rather the result of local kinematics, e.g., 
an outflow from a galaxy or clump within the nebula.  
It is possible that PRG1 resembles a scaled up version of the ``clumpy disks'' 
seen at $z\approx2$ \citep[e.g.,][]{forster2009,newman2012}, 
which in some cases show a peak in the velocity dispersion that is offset by 
several kiloparsecs from the disk center 
due to the presence of a large star-forming clump driving an outflow.  In the case of PRG1, this 
possibility is supported by the presence of a continuum source near the location of the peak velocity 
dispersion that is visible in recently acquired $HST$/WFC3 F140W imaging (Prescott et al., in prep.).

\begin{figure}[h]
\includegraphics[angle=0,width=5.5in]{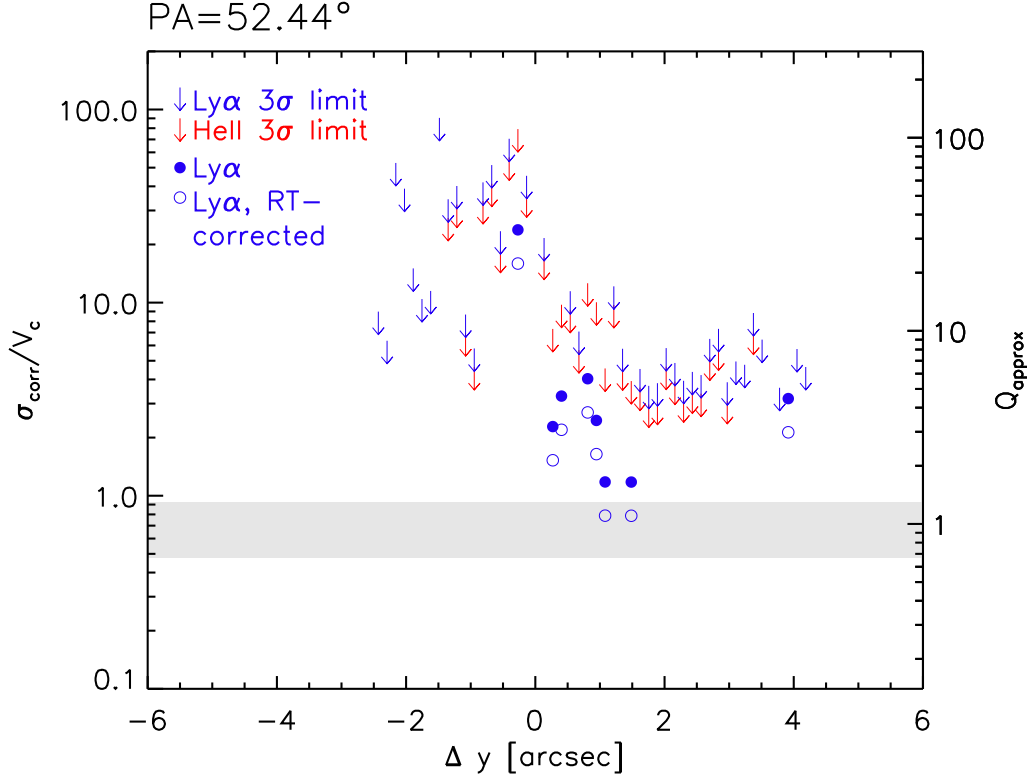} 
\caption[]{
Measured $\sigma_{corr}/V_{c}$ as a function of position along the PA=52\fdg44 slit, which 
coincides roughly the major axis of the proposed disk.  
Upper limits (3$\sigma$) are plotted for apertures where the measured linewidth is consistent 
with the instrumental resolution; otherwise the $\sigma_{corr}$ values are used, i.e., corrected for 
the instrumental resolution.  $V_{c}$ is the measured velocity from the \heii\ line.  
The \lya\ measurements are shown as solid circles, while the open blue circles represent the result of 
applying an approximate ``radiative transfer correction,'' i.e., scaling the \lya\ $\sigma_{corr}$ measurements 
such that the \lya\ and \heii\ values agree in the one aperture along the PA=146\fdg0 slit where both lines 
are clearly resolved.  The right axis gives the corresponding approximate Toomre Q values, 
under the assumption of a marginally stable disk.  
In this case, $Q_{approx}\approx (a/1.4) \times (1.0/f_{gas}) \times \sigma_{corr}/V_{c}$, following 
\citet{genzel2014}, with $a$ being a geometric factor that can take values of [1,1.4,2] for a 
Keplerian rotation curve, a flat rotation curve, and a solid-body rotation curve, respectively, 
and $f_{gas}$ being the gas mass fraction.  The grey-shading corresponds to the approximate values 
of $Q$ below which the gas is expected to be unstable to collapse. 
}
\label{fig:sigvelo}
\end{figure}

Despite its simplicity, the toy model provides a good representation of the velocity profiles observed in PRG1.  
To estimate the stability of the proposed disk, 
we plot the ratio of the velocity dispersion (corrected for the instrumental resolution, $\sigma_{corr}$) and the 
circular velocity ($V_{c}$, measured from \heii) as a function of position along the PA=52\fdg44 slit (Figure~\ref{fig:sigvelo}); 
as the PA=52\fdg44 slit corresponds roughly to the major axis of the disk and the preferred disk inclination 
is relatively high, we do not apply any corrections for the disk inclination or azimuthal angle within the disk, i.e., 
we take $V_{c}\approx V_{obs}$, the observed velocity.  
Apertures where the measured linewidth is consistent with the instrumental resolution are shown 
as upper limits (3$\sigma$).  Using the one aperture along PA=146\fdg0 where both \lya\ and \heii\ 
are clearly resolved, we compute an approximate ``radiative transfer correction'' for \lya, i.e., 
the factor by which the \lya\ linewidth should be scaled down in order to match that of the non-resonant 
\heii\ line.  Under the crude assumption that this factor can be applied across the entire nebula, 
this approach provides a means of peering below our instrumental resolution limit.  
The \lya\ measurements are then plotted both with and without this ``radiative transfer correction.''  
Following \citet{genzel2014} and assuming a marginally stable disk, we estimate the approximate Toomre Q 
parameter, a measure of whether the gas will be unstable to collapse and result in subsequent star formation, where 
$Q_{approx}\approx(a/1.4)\times(1.0/f_{gas})\times\sigma_{corr}/V_{c}$, with $a$ being a geometric factor with values of 
[1.0,1.4,2.0] for a Keplerian rotation curve, a flat rotation curve, and a solid-body rotation curve, respectively, 
and with $f_{gas}$ being the gas mass fraction.  Over most of the nebula, we can only report upper limits 
on the $\sigma_{corr}/V_{c}$, but in regions where 
the lines are resolved, we estimate that $Q_{approx}$ is typically greater than 0.67-1.3, i.e., the critical values 
below which the gas becomes unstable, even under the assumption of an extremely high gas fraction ($f_{gas}=1$).
Thus in most of the apertures where we resolve the \lya\ emission line, the nebula appears 
to be stable against collapse.  In a few central apertures as well as at larger radii, however, the gas may 
be unstable to collapse and subsequent star formation.  

Deep IFU observations of \lya\ nebula such as PRG1 detecting multiple emission lines will be important for shedding further 
light on the complex kinematics of these systems, but the suggestion from the data presented here is that the gas in PRG1 
is undergoing large-scale rotation in a clumpy, turbulent disk.

\subsection{Implications}

Recent high resolution numerical simulations of Milky Way mass halos 
\citep[$\sim10^{12}$ M$_{\odot}$ at $z=0$, which corresponds to $\sim10^{11}$ M$_{\odot}$ at $z=3$, 
assuming the halo growth rate from][]{neistein06} 
have indicated that newly accreted gas will have high angular momentum, spending 1-2 dynamical times 
in the outer halo as a ``cold flow disk" that extends to many tens of kiloparsecs outside the 
central galaxy \citep{stewart11,stewart13}.  Although the halo masses of \lya\ nebulae are 
poorly constrained, it is possible that we are seeing a similar phenomenon in \lya\ nebulae 
systems like PRG1 and SSA22-LAB2 \citep{martinchris14}, i.e., the early formation of a large 
Milky Way mass galaxy or galaxy group.  In the case of PRG1, the rotation period implied 
by our disk modeling ($T_{rot}\approx4.9\times10^{8}$ yr, assuming $V_{max}\approx250$ km s$^{-1}$ 
at $R=20$~kpc) is consistent with the system undergoing a handful of rotations by $z\approx1.67$ 
(when the age of the Universe was $\sim$3.9 Gyr). 
Our results also motivate further high resolution theoretical work on the angular 
momentum of cold gas accretion as a function of halo mass, particularly for the higher mass halos 
thought to host giant \lya\ nebulae \citep[e.g.,][]{pres08,yang09,yang10}.

At the same time, our data do not favor the idea that gravitational cooling 
is the dominant {\it powering} mechanism responsible for the \lya\ emission in this system.  
Gravitational infall of lower metallicity gas (i.e., ``cold flows'') would not be expected to produce such strong 
\heii\ emission over such a large spatial extent \citep{yang06,rosdahl12}, and the presence of \civ\ and \ciii\ 
indicates the gas is at least somewhat enriched.  In addition, cold flow powered \lya\ nebulae are predicted to 
exhibit \lya\ emission line profiles with a dominant blue peak, owing to infall \citep{faucher10}, whereas in 
PRG1 we find that whether \lya\ is observed to be redshifted or blueshifted relative to \heii\ depends on the 
position within the nebula, with most locations showing redshifted \lya\ emission.  
In addition, there is still debate as to whether gravitational cooling can provide the \lya\ luminosities that are 
typically observed in \lya\ nebulae \citep[e.g.,][]{goerdt10,faucher10,rosdahl12}.

What does seem likely is that the gas reservoir in these regions 
is being supplied by recent accretion, perhaps coming in with significant angular momentum, but 
with the gas being illuminated and photoionized by a powerful source of ionizing photons, 
i.e., highly obscured AGN and star formation that is being fueled by the same accretion event.  
In PRG1, this scenario would lead to the observed strong, spatially extended \lya, \heii, and 
metal line emission, and to the small velocity offset -- primarily to the red -- that is measured for 
\lya.\footnote{In the context of AGN fluorescence, the presence of both blueshifted and redshift \lya\ is easily understood, since the \lya\ line is expected to exhibit either a prominent red or blue peak depending on the geometric alignment 
of the AGN with respect to the gas velocity field \citep{cantalupo05}.} 
This scenario would also be consistent with the observation of several \lya\ nebulae that appear 
to be aligned with the filament of galaxies they reside in \citep{erb11}.  One can imagine that 
the alignment is due to the preferential flow of material within the filament, perhaps entering 
a messy, rotating disk as it feeds a growing galaxy or protocluster.  Detailed kinematic 
studies using high spatial resolution IFU observations to look for evidence 
of rotation or coherent flows within a larger number of \lya\ nebulae 
would be ideal for testing this hypothesis.

\section{Conclusions}
\label{sec:conclusions}

Using the spatially extended emission in \lya\ as well as in less optically thick 
emission lines, we study 
the role of scattering and the kinematics of the extended gas within PRG1, a \lya\ nebula at $z\approx1.67$.
The low measured kinematic offset of \lya\ and the similarity of the surface brightness profiles observed in 
different emission lines are strong arguments that the extended \lya\ is being produced in situ within the 
spatially extended gas, most likely due to photoionization from 
an AGN or distributed star formation, rather than scattered from a central source.  
The large-scale coherent velocity shear we observe across the entire nebula -- 
$\approx$500 km~s$^{-1}$ over the central $\approx$50~kpc -- is broadly consistent with 
large-scale rotation in a clumpy, turbulent disk that is at least partially stable against collapse.  
Thus, while our data are inconsistent with cooling radiation powering the \lya\ emission, 
accreting gas with high angular momentum flowing in along cold streams may explain the 
large scale coherent velocity structure 
that we observed within the extended \lya\ nebula.  
This work suggests that, in at least some cases, the resonant \lya\ line 
can be a robust tracer of the large-scale kinematics, 
and it motivates further deep spectroscopic studies of the extended gas within \lya\ nebulae 
as a probe of the kinematics of the gas reservoir fueling episodes of active 
galaxy formation.


\acknowledgments
The authors thank Mark Dijkstra, Kristian Finlator, Peter Laursen, 
Norm Murray, and Anna Pancoast for illuminating 
discussions, Kristian Finlator for observing assistance, 
Alice Shapley for providing a comparison sky spectrum used to check 
the accuracy of our wavelength calibration, and the anonymous referee 
for useful suggestions that improved the quality of this paper. 
M.K.M.P. was supported by a TABASGO Prize Postdoctoral Fellowship and 
a Dark Cosmology Centre Postdoctoral Fellowship.  This research was also supported 
in part by the National Science Foundation under AST-1109288 (C.L.M.), 
and by NOAO (A.D.).  NOAO is operated by the Association of Universities 
for Research in Astronomy (AURA), Inc. under a cooperative agreement with 
the National Science Foundation.  AD's research is also partially supported 
by the Radcliffe Institute for Advanced Study and the Institute for Theory 
and Computation at Harvard University. AD thanks the Aspen Center for Physics, 
which is supported by the National Science Foundation Grant No. PHY-1066293.

The data presented herein were obtained at the W.M. Keck Observatory, which is operated as a 
scientific partnership among the California Institute of Technology, the University of California, 
and the National Aeronautics and Space Administration.  The Observatory was made possible by the 
generous financial support of the W.M. Keck Foundation.  The authors wish to recognize and acknowledge 
the very significant cultural role and reverence that the summit of Mauna Kea has always had within the 
indigenous Hawaiian community.  We are most fortunate to have the opportunity to conduct observations 
from this mountain. 



\clearpage

\begin{deluxetable}{cccccccccc}
\tabletypesize{\scriptsize}
\tablecaption{Observing Log} 
\tablewidth{0pt}
\tablehead{
\colhead{UT Date} & \colhead{UT Time} & \colhead{Parallactic} & \colhead{Right Ascension\tablenotemark{a}} & \colhead{Declination\tablenotemark{a}} 
& \colhead{Position} & \colhead{Exposure\tablenotemark{b}} & \colhead{Slitwidth} & \colhead{Seeing} & \colhead{Conditions} \\ 
 &  Range & Angle Range & (hours) & (degrees) & Angle  &  Time (sec) & (arcsec) & (arcsec) & } 
\startdata
2009 April 25 & 10:00-12:45 & -145\degree-111\degree\ & 14:35:12.385 & 35:11:06.62 & 52.44\degree\ &  8300 / 8100 & 1.2 & 0.8-1.0 & Clear \\ 
2009 April 25 & 8:00-10:00, & 78\degree-35\degree, & 14:35:12.385 & 35:11:06.62 & 146.0\degree\ &  7630 / 7200 & 1.2 & 1.0-1.1 & Clear \\ 
& 13:00-14:30 & 108\degree-91\degree\ & & & &  & & & \\ 
2010 May 10   & 6:45-10:00 & -99\degree-167\degree\  & 14:35:12.385 &  35:11:06.62 & 52.44\degree\ &  10350 / 9000 & 1.5 & 0.8-0.9 & Clear \\ 
2010 May 10   & 10:45-14:15 & 135\degree-85\degree\ & 14:35:12.385 &  35:11:06.62 & 146.0\degree\ &  10350 / 9900 & 1.5 & 0.7-0.8 & Clear \\ 
\enddata
\tablenotetext{a}{Location on the nebula where the two slits cross.}
\tablenotetext{b}{Values listed separately for LRIS-Blue / LRIS-Red.}
\label{tab:specobs}
\end{deluxetable}

\clearpage

\begin{deluxetable}{cccccccc}
\tabletypesize{\scriptsize}
\tablecaption{Spectroscopic Calibration} 
\tablewidth{0pt}
\tablehead{
\colhead{UT Date} & \colhead{Position} & \colhead{Detector} & \colhead{Dispersion} & 
\colhead{Resolution FWHM} & 
\colhead{Wavelength} & \colhead{Wavelength Solution} & \colhead{Sky Line Comparison\tablenotemark{a}} \\ 
 & Angle  & & (\AA/pix)  & (km s$^{-1}$) & Range (\AA) & rms (\AA) & rms (\AA) } 
\startdata

2009 April 25 & 52.44\degree\ & LRIS-Blue & 1.09  & 758-501 & 3100-5600 & 0.24 & 0.32 \\ 
		& 		& LRIS-Red & 1.86 & 471-350 & 5600-10000 & 0.11 & 0.20 \\ 
2009 April 25 & 146.0\degree\ & LRIS-Blue & 1.09  & 758-501 & 3100-5600 & 0.24 & 0.25 \\ 
		& 		& LRIS-Red & 1.86 & 471-350 & 5600-10000 & 0.12 & 0.42 \\ 
2010 May 10   & 52.44\degree\ & LRIS-Blue & 2.18  & 966-554 & 3100-6800 & 0.14 & 0.14 \\ 
		& 		& LRIS-Red & 1.16 & 439-292 & 6800-10870 & 0.21 & 0.25  \\ 
2010 May 10   & 146.0\degree\ & LRIS-Blue & 2.18  & 966-554 & 3100-6800 & 0.16 & 0.17 \\ 
		& 		& LRIS-Red & 1.16 & 439-292 & 6800-10870 & 0.28 & 0.42 \\ 

\enddata
\tablenotetext{a}{Measured rms relative to sky lines HgI$\lambda4047$, HgI$\lambda5461$ (2009), \oi$\lambda5577$, and \oi$\lambda6300$ (2010) for LRIS-Blue and \oi$\lambda6364$, OH8-3(P1,2)$\lambda7316$, OH7-3(P1,2)$\lambda8886$, and OH0-3(P1,2)$\lambda9872$ (2009) or OH9-5(P1,2)$\lambda10082$ (2010) for LRIS-Red.}
\label{tab:redux}
\end{deluxetable}

\begin{deluxetable}{cccccccc}
\tabletypesize{\scriptsize}
\tablecaption{Surface Brightness Profile Sizes}
\tablewidth{0pt}
\tablehead{
\colhead{PA} & \colhead{Line} & \colhead{D$_{50}$\tablenotemark{a}} & \colhead{D$_{50}$} & \colhead{F$_{50}$\tablenotemark{b}} & \colhead{D$_{90}$\tablenotemark{a}} & \colhead{D$_{90}$} & \colhead{F$_{90}$\tablenotemark{b}} \\
         &     & (arcsec) & (kpc) & (10$^{-17}$ erg s$^{-1}$ cm$^{-2}$) & (arcsec) & (kpc) & (10$^{-17}$ erg s$^{-1}$ cm$^{-2}$) }
\startdata
 & & & \\
       52\fdg44 &           Ly$\alpha$ &       3.05$\pm$      0.11 &      25.87$\pm$      0.94 &       60.5$\pm$       2.0 &       7.14$\pm$      0.37 &      60.45$\pm$      3.14 &      108.9$\pm$       3.0\\
                &                 HeII &       2.29$\pm$      0.16 &      19.43$\pm$      1.38 &        5.1$\pm$       0.3 &       6.34$\pm$      0.84 &      53.67$\pm$      7.11 &        9.1$\pm$       0.5\\
                &       CIV+HeII+CIII] &       2.36$\pm$      0.13 &      19.96$\pm$      1.09 &       11.3$\pm$       0.6 &       5.64$\pm$      0.67 &      47.73$\pm$      5.64 &       20.3$\pm$       1.0\\
 & & & \\
\hline
 & & & \\
       146\fdg0 &           Ly$\alpha$ &       2.29$\pm$      0.13 &      19.39$\pm$      1.08 &       43.8$\pm$       2.0 &       7.00$\pm$      0.60 &      59.32$\pm$      5.10 &       79.1$\pm$       3.4\\
                &                 HeII &       1.91$\pm$      0.17 &      16.19$\pm$      1.43 &        3.9$\pm$       0.3 &       4.79$\pm$      0.59 &      40.53$\pm$      5.02 &        7.0$\pm$       0.5\\
                &       CIV+HeII+CIII] &       1.72$\pm$      0.12 &      14.54$\pm$      1.01 &        7.5$\pm$       0.5 &       4.25$\pm$      0.38 &      36.01$\pm$      3.22 &       13.5$\pm$       0.8\\
\enddata
\tablenotetext{a}{Diameter containing 50\% or 90\% of the total flux measured in a given emission line.}
\tablenotetext{b}{Total flux contained within D$_{50}$ or D$_{90}$ in a given emission line.}
\label{tab:sbmodel}
\end{deluxetable}

\clearpage

\begin{deluxetable}{ccc}
\tabletypesize{\scriptsize}
\tablecaption{Toy Model Fit Parameters}
\tablewidth{0pt}
\tablehead{
\colhead{Parameter} & \colhead{67\% Confidence}  \\
         &  Interval }
\startdata
 $\Theta_{off}$ & [       -30$^\circ$,       -16$^\circ$] \\
            $i$ & [        38$^\circ$,        77$^\circ$] \\
      $R_{max}$ & [       0.8$\arcsec$,       2.3$\arcsec$] \\
      $V_{max}$ & [       233,       441]\\
      $X_{c}$ & [      -0.9$\arcsec$,      -0.3$\arcsec$] \\
      $Y_{c}$ & [      -0.7$\arcsec$,      -0.4$\arcsec$] \\
\enddata
\label{tab:toymodel}
\end{deluxetable}


\clearpage

\clearpage

\clearpage

\clearpage 

\clearpage

\clearpage

\clearpage

\clearpage

\clearpage

\clearpage

\clearpage

\clearpage

\clearpage

\clearpage

\clearpage

\clearpage
\end{document}